\documentclass{aa}
\usepackage{graphicx}
\usepackage{txfonts}
\usepackage{hyperref}
\hypersetup{colorlinks = true, citecolor = {blue}, urlcolor= {blue}}

\def\PGPU{$\varphi-$GPU }

\def\gapprox{\;\rlap{\lower 3.0pt                       
        \hbox{$\sim$}}\raise 2.5pt\hbox{$>$}\;}
\def\lapprox{\;\rlap{\lower 3.1pt                       
        \hbox{$\sim$}}\raise 2.7pt\hbox{$<$}\;}

\newcommand{\be}{ \begin{equation} }
\newcommand{\ee}{\end{equation}}

\newcommand{\ben}{\begin{enumerate}}
\newcommand{\een}{\end{enumerate}}

\renewenvironment{thebibliography}[1]{\begin{oldthebibliography}{#1}\setlength{\parskip}{0ex}\setlength{\itemsep}{0ex}}{\end{oldthebibliography}}

\usepackage{diagbox}
\usepackage{siunitx}
\newcommand{\orcid}[1]{\href{https://orcid.org/#1}{\protect\includegraphics[width=8pt]{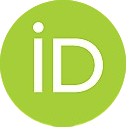}}}
\makeatletter
\renewcommand*\aa@pageof{, page \thepage{} of \pageref*{LastPage}}
\makeatother

\usepackage[dvipsnames]{xcolor}

\definecolor{darkgreen}{RGB}{31, 207, 31}

\begin{document}

\title{Dynamical evolution of Milky Way globular clusters on the cosmological timescale\\  I. Mass loss and interaction with the nuclear star cluster}


\author{Maryna~Ishchenko
\inst{1,2,3}\orcid{0000-0002-6961-8170}
\and
Peter~Berczik
\inst{2,3,8,1}\orcid{0000-0003-4176-152X}
\and
Taras~Panamarev
\inst{7,3}\orcid{0000-0002-1090-4463}
\and
Dana~Kuvatova
\inst{3}\orcid{0000-0002-5937-4985}
\and
Mukhagali~Kalambay
\inst{3,4,5,6}\orcid{0000-0002-0570-7270}
\and
Anton~Gluchshenko
\inst{3}\orcid{0000-0002-0738-7725}
\and
Oleksandr~Veles
\inst{1,2}\orcid{0000-0001-5221-2513}
\and
Margaryta~Sobolenko
\inst{1,2}\orcid{0000-0003-0553-7301}
\and
Olexander Sobodar
\inst{1,2}\orcid{0000-0001-5788-9996}
\and
Chingis~Omarov
\inst{3}\orcid{0000-0002-1672-894X}
}

\institute{Main Astronomical Observatory, National 
           Academy of Sciences of Ukraine,
           27 Akademika Zabolotnoho St, 03143 Kyiv, Ukraine  \email{\href{mailto:marina@mao.kiev.ua}{marina@mao.kiev.ua}}
           \and
Nicolaus Copernicus Astronomical Centre Polish Academy of Sciences, ul. Bartycka 18, 00-716 Warsaw, Poland
\and
           Fesenkov Astrophysical Institute, Observatory 23, 050020 Almaty, Kazakhstan
           \and
           Heriot-Watt International Faculty, K. Zhubanov Aktobe Regional University, Zhubanov broth. str. 263, 030000 Aktobe, Kazakhstan
           \and
           Energetic Cosmos Laboratory, Nazarbayev University, Kabanbay Batyr ave. 53, 010000 Astana, Kazakhstan
           \and
           Faculty of Physics and Technology, Al-Farabi Kazakh National University,
           al-Farabi ave. 71, 050040 Almaty, Kazakhstan
           \and 
           Rudolf Peierls Centre for Theoretical Physics, Parks Road, OX1 3PU, Oxford, UK
           \and
           Konkoly Observatory, Research Centre for Astronomy and Earth Sciences, HUN-REN CSFK, MTA Centre of Excellence, Konkoly Thege Mikl\'os \'ut 15-17, 1121 Budapest, Hungary
           }
   
\date{Received xxx / Accepted xxx}

\abstract
{Based on the \textit{Gaia} DR3, we reconstructed the orbital evolution of the known Milky Way globular clusters and found that six objects, NGC~6681, NGC~6981, Palomar 6, NGC~6642, HP~1, and NGC~1904, very likely interact closely with the nuclear star cluster.}
{We study the dynamical evolution of selected Milky Way globular clusters and their interactions with the Galactic centre over cosmological timescales. We examine the global dynamical mass loss of these globular cluster systems, their close interactions with the Galactic centre, and the potential capture of stars by the Milky Way nuclear star cluster.}
{For the dynamical modelling of the clusters, we used the parallel $N$-body code $\varphi$-GPU, which allows star-by-star simulations of the systems. Our current code also enabled us to follow the stellar evolution of individual particles, including the formation of high-mass remnants. The modelling was carried out in a Milky Way-like, time-variable potential (with a dynamically changing mass and scale length), obtained from the IllustrisTNG-100 database, with a full integration time of eight billion years.}
{Based on extensive numerical modelling and analysis, we estimated the mass loss and the global and inner structures of the selected six clusters. Over an evolution of eight billion years, the clusters lost $\approx$80\% of their initial mass. We analysed the phase-space evolution of the individual unbound stars NGC~6681, NGC~6642, HP 1, and NGC~1904. We found that only NGC~6642 could potentially have been a source for populating the Milky Way nuclear star cluster in the past.}
{}

\keywords{Galaxy: globular clusters: general - globular clusters: individual: NGC~6681, NGC~6981, Palomar 6, NGC~6642, HP 1, NGC~1904 - Galaxy: center - Methods: numerical}

\titlerunning{GCs mass loss and interaction with nuclear star cluster}
\authorrunning{M.~Ishchenko et al.}
\maketitle

\section{Introduction}\label{sec:Intr}

Globular clusters (GCs) are among the oldest stellar systems in the Universe \citep{Vandenberg1996}. These tightly bound collections of hundreds of thousands and sometimes millions of stars offer valuable insights into galaxy formation and evolution \citep{Harris1991, Brodie2006}. Understanding the interplay between GCs and the central regions of galaxies, particularly those harbouring nuclear star clusters (NSCs) and supermassive black holes (SMBHs), has important implications for models of Galactic formation \citep{Bland-Hawthorn2016, Neumayer2020}.

The formation of NSCs in galaxies such as the Milky Way (MW) and others, including M~31, has been linked to the inspiral of GCs due to dynamical friction, as initially proposed by \citet{Tremaine1975}. Observational evidence supporting this assertion includes the deficit of massive GCs within the inner regions of galaxies
\citep{Lotz2001, Capuzzo-Dolcetta2009}. 

Simulations have shown that dynamical friction can effectively form NSCs within a Hubble time under certain conditions \citep{Capuzzo-Dolcetta1993, Oh2000, Lotz2001, Agarwal2011, Neumayer2011}. However, \citet{Hartmann2011} and \citet{Antonini2012}, for example, suggested that only about half of an NSC mass may come from this process, with the rest possibly originating from accreted gas and in situ star formation. Additionally, the growth of NSCs can be facilitated by capturing stars from GCs that pass close by, but that ultimately survive. Our recent study \citep[][hereafter \hyperlink{I23}{\color{blue}{Paper~I}}]{Ishchenko2023b} identified ten GCs in the MW that have approached the Galactic centre (GalC) to within a distance of 100~pc, illustrating this mechanism of NSC growth.

The properties of the NSC of the MW are critical for understanding its formation and evolutionary history \citep[see][for a review]{Neumayer2020}. Its effective radius falls within the range of $r_\mathrm{eff}$ between $4.2\pm0.4$~pc and $7.2\pm2.0$~pc \citep{Schodel2014a, Fritz2016}. The total stellar mass of the MW NSC is estimated to be between $(2.1\pm0.7)\times 10^{7}\rm\;M_\odot$ and $(4.2\pm1.1)\times10^{7}\rm\;M_\odot$ with photometric and dynamical methods \citep{Schodel2014a, Fritz2016, Feldmeier-Krause2017}. Furthermore, the star formation history within the NSC indicates that more than 80\% of its stellar mass was formed over five billion years ago, initially characterised by a high star formation rate that diminished to a minimum between one and two billion years ago \citep{Blum2003, Pfuhl2011, Nogueras-Lara2020}.

Our recent investigation \hyperlink{I23}{\color{blue}{Paper~I}} was dedicated to finding the MW GCs that might interact with the NSC and the SMBH during their lifetime. For this purpose, we applied several time-variable external potentials from the cosmological IllustrisTNG-100 database (TNG-TVP) to be more physically motivated in our simulations \citep[more information about selected potentials can be found in][hereafter \hyperlink{I23a}{\color{blue}{Paper~II}}]{Ishchenko2023a}. Taking into account the errors of the measurements from \textit{Gaia}~DR3 catalogue, we carried out 1000~simulations for each GC from the catalogue and integrated GC orbits up to ten billion years in lookback time. Applying criteria in 100~pc for the relative distance between GC and GalC fly-by, we found ten GCs. The six GCs NGC~6401, Palomar~6, NGC~6681, NGC~6712, NGC~6287, and NGC~6642 have very reliable close passages near the GalC in all our external potentials with a probability of almost 100\%. The other four GCs, NGC~6981, HP~1, NGC~1904, and NGC~362, have interactions probability values in the range from 90\% to 27\%. 

The main idea of this work is to carry out a full dynamical $N$-body modelling of the selected GCs that also includes the detail stellar evolution of the selected objects. We analyse the global GC evolution and the mass distribution in the inner structure,  including the compact stellar remnants. We also estimate the potential stellar population loss from the GCs that is due to the influence of the GalC with the SMBH and the rate of the stellar capturing by the NSC in a time frame of eight billion years. 

The paper is organised as follows. In Section~\ref{sec:init-integr} we present the initial conditions for the selected GCs and for the integration procedure. We present the inner and global stellar evolution of the high remnants in Section~\ref{sec:inn-glob-evol}. In Section \ref{sec:int-nsc} we present the probability analysis of the stellar accumulation onto the Galactic central NSC. In Section \ref{sec:disc-con} we discuss and summarise our results.

In Appendix~\ref{app:gc-lit} we present the overview of the GCs we selected that might produce the stellar population in the central part of the MW. In Appendix~\ref{app:NS-distr} we present the current neutron star distribution in the Galaxy. 

\section{Initial conditions and integration procedure for globular clusters}\label{sec:init-integr}

\subsection{Selection procedure for the globular clusters}\label{subsec:extra-sel}

In our previous work, \hyperlink{I23}{\color{blue}{Paper~I}}, we modelled GCs as point-mass particles in time-varying potentials. We identified ten clusters that had close encounters with the GalC. We pre-selected our GCs with the following criterion: In order   to detect the event of  such a close passage, the relative distance between the GalC and the GC should be shorter than 100 pc (see Table 3 in \hyperlink{I23}{\color{blue}{Paper~I}}). To be more robust in our pre-selected list, we performed 1000 simulations in which we varied the initial velocities of the GCs within $\pm1\sigma$ of the measurement errors taken from the normal distribution \citep{Baumgardt2021}. 

Here, we aim to study the mass loss from the GCs and the potential star captures by the NSC. This requires a full direct $N$-body modelling of the clusters and means that the point-mass approximation is no longer valid. However, given the computational intensity of direct $N$-body simulations, we need to refine the selection of the GCs.

\begin{figure}[htbp!]
\centering
\includegraphics[width=0.99\linewidth]{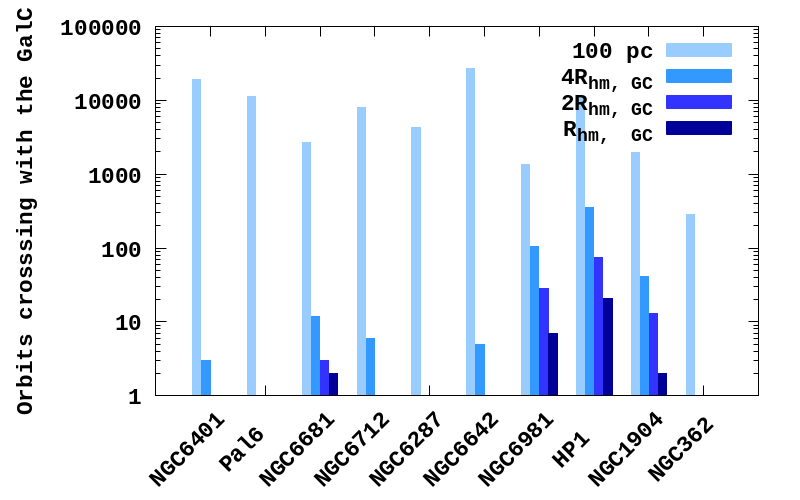}
\caption{Orbits crossings with the GalC for selected GCs for the four different distance criteria in {\tt 411321} TNG-TVP external potential. Dark blue corresponds to the r$_{\rm hm}$ distance from GalC, and light blue shows the distance of 100 pc.}
\label{fig:gc-coll}
\end{figure}

\paragraph{Additional distance criteria.}  To evaluate the frequency of close encounters between GCs and the GalC, we implemented several distance criteria to quantify their proximity: $N_{100\rm}$, $N_{4r_{\rm hm}}$, $N_{2r_{\rm hm}}$, and $N_{r_{\rm hm}}$. Here, $N_{100\rm}$ specifies instances in which the separation between a GC and the GalC is smaller than 100~pc; $N_{4r_{\rm hm}}$ and $N_{2r_{\rm hm}}$ indicate distances smaller than four and two times the sum of the half-mass radii of the GCs, respectively; and $N_{r_{\rm hm}}$ refers to distances smaller than the half-mass radius of the GCs. The present-day half-mass radii ($r_{\rm hm}$) for each GC are listed in Table~\ref{tab:phis-gc}. We assessed the potential for close passages by applying these four distinct distance criteria, as illustrated in Fig.~\ref{fig:gc-coll}. The figure visually depicts the criteria we applied to the ten GCs in the {\tt 411321} TNG-TVP external potential (\hyperlink{I23}{\color{blue}{Paper~I}}, \hyperlink{I23a}{\color{blue}{Paper~II}}).

Fig.~\ref{fig:gc-coll} reveals that several GCs experience very close approaches to the GalC. This matches the criterion of $N_{r_{\rm hm}}$ and indicates the possibility for collisions with both the SMBH and the NSC. This is highlighted in dark blue. It is crucial to emphasise that the likelihood of these collisions is generally low, estimated at a few percent, and a collision might occur once or twice throughout the orbital evolution of the GCs. However, for specific clusters, namely for NGC~6681, NGC~6981, NGC~6712, NGC~6642, HP1, and NGC~1904, the analysis in the {\tt 411321} TNG-TVP potential indicates a noteworthy persistence of these collision chances.

\begin{figure}[htbp!]
\centering
\includegraphics[width=0.9\linewidth]{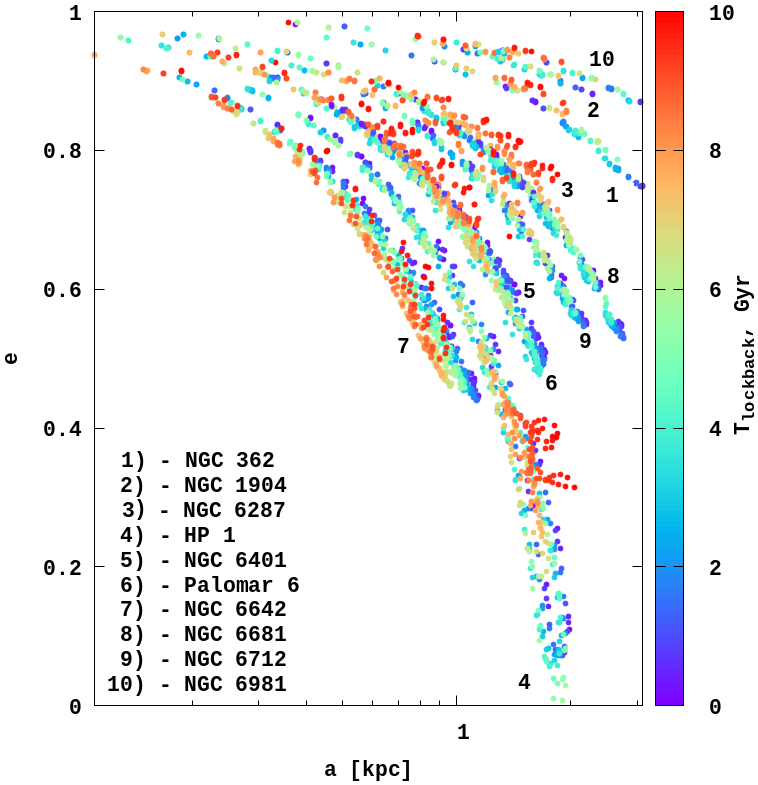}
\caption{Evolution of the semimajor axis and eccentricity during the whole forward-integration time for selected GCs in the {\tt 411321} TNG-TVP potential. Time is represented by the colour-codding.}
\label{fig:gc-e-a}
\end{figure}

\paragraph{Evolution of orbital parameters and orbit types.} To illustrate the temporal evolution of the orbital elements for the ten GCs under study, we present data within the {\tt 411321} TNG-TVP potential in Fig.~\ref{fig:gc-e-a}. This visualisation highlights the progression of eccentricities from higher to lower values. We classified the GCs into three distinct categories based on the behaviour of their eccentricity $e$.  The first category includes GCs such as NGC~6981, NGC~1904, and NGC~362, which exhibit a relatively narrow eccentricity range that is consistently between 0.8 and 1 throughout their orbital evolution. The second category comprises GCs including NGC~6681, NGC~6712, Palomar~6, NGC~6287, NGC~6401, and NGC~6642, which are characterised by a much broader variation in eccentricity ($e$: 0.4--1). The most notable case is HP~1, which is classified into the third category. Its eccentricity varies significantly and transitions from one to nearly zero. This indicates a shift from a highly elliptical to a nearly circular orbit during the lifetime of the cluster.

We further refined our analysis by considering the relative orbital fly-by velocities and types of orbits. GCs NGC~6401, NGC~6642, Palomar~6, and NGC~6681 exhibit relative fly-by velocities between 262 and 410 km~s$^{-1}$, indicating potentially close passages near the GalC (Table \ref{tab:phis-gc}, column (3)), which were discussed in \hyperlink{I23}{\color{blue}{Paper~I}}. Three of these follow tube-like orbits, while NGC~6681 features a more radial trajectory. Their orbital patterns suggest that these GCs originated within the MW bulge; NGC~6681 is a notable exception. Particular emphasis is placed on NGC~6981, which is characterised by a narrow eccentricity range, a long radial orbit extending to 15~kpc, the highest velocity of our selection, and a unique ex situ origin. Finally, we include HP~1 as a contrasting case study. HP~1, the oldest GC in the MW, has a complex, irregular orbit with wide eccentricity fluctuations, and similar to the other bulge-region GCs, it likely originated within our own Galaxy.

As a result, we selected six clusters for further direct $N$-body modelling: NGC~6401, NGC~6642, Palomar~6, NGC~6681, NGC~6981, and HP~1. Table~\ref{tab:phis-gc} presents their key characteristics, with a focus on current masses and ${r_{\rm hm}}$, primarily derived from \cite{Baumgardt2021}. We also include data on relative velocities and orbital types (TO) from \hyperlink{I23a}{\color{blue}{Paper~II}}, as well as possible progenitors (Pg.) based on \cite{Malhan2022}, where M-B denotes the MW main bulge, G-E signifies Gaia-Enceladus remnants, and Pontus represents an ancient satellite galaxy. Finally, individual metallicity $Z_{\rm m}$ values for the GCs are included based on \cite{Boyles2011}.

\begin{table}[htbp]
\caption{Main kinematic and physical characteristics of the selected GCs at present.}
\centering
\sisetup{separate-uncertainty}
\resizebox{0.50\textwidth}{!}{
\begin{tabular}{
c
S[table-format=2.2] 
cc
S[table-format=2.2] 
S[table-format=2.1] 
cc
S[table-format=2.1] 
c}
\hline
\hline
GC & {$M$} & {$dV$} & Age & $r_{\rm hm}$ & $Z_{\rm m}$ & TO & Pg. \\
 & {$10^{4}\rm\;M_{\odot}$} & km/s & Gyr & pc & {$10^{-4}$} &  &  \\
(1) & {(2)} & (3) & (4) & {(5)} & {(6)} & {(7)} & {(8)}\\
\hline
\hline
NGC~6401  & 12.10 & 331 & 13.20$^{\rm a}$ & 3.17 & 0.85 & TB & M-B \\
Palomar 6     &  8.60 & 340 & 12.40$^{\rm b}$ & 2.91 & 2.46 & TB & M-B \\
NGC~6681  & 10.50 & 410 & 12.80$^{\rm c}$ & 3.04 & 0.46 & LR & --  \\
NGC~6642  &  3.90 & 262 & 13.80$^{\rm d}$ & 1.73 & 1.10 & TB & M-B \\
NGC~6981  &  8.10 & 545 & 10.88$^{\rm c}$ & 5.79 & 0.66 & LR & G-E \\
HP~1      & 14.00 & 304 & 12.80$^{\rm e}$ & 4.19 & 2.00 & IR & M-B \\
\hline
\end{tabular}
}
\tablefoot{
Masses and $r_{\rm hm}$ in columns (2, 3) is up to Oct. 2023 \url{https://people.smp.uq.edu.au/HolgerBaumgardt/globular/};
Column (4) -- the GCs age in Gyr according to the: 
$^{\rm a}$\cite{Cohen2021}
$^{\rm b}$\cite{Souza2021},
$^{\rm c}$\cite{Forbes2010},
$^{\rm d}$\cite{Balbinot2009},
$^{\rm e}$\cite{Ortolani2011};
}
\label{tab:phis-gc}
\end{table} 

\subsection{Initial conditions and integration}\label{subsec:init-cond}

For the dynamical orbital integration of GCs, including the effects of stellar evolution, we employed the high-order parallel $N$-body code \PGPU\footnote{$N$-body code \PGPU: \\~\url{ https://github.com/berczik/phi-GPU-mole}}, which is based on the fourth-order Hermite integration scheme with hierarchical individual block time steps \citep{Berczik2011,BSW2013}. The current version of the \PGPU code uses native GPU support and direct code access to the GPU using the NVIDIA native CUDA library. This code is well tested and has already been used to obtain important results in our prior GC simulations \citep{Just2009, Shukirgaliyev2017, Shukirgaliyev2018, Shukirgaliyev2021}. 

The current version of the code also incorporates the most recent stellar evolution models. The most important updates were made to components of stellar evolution such as an updated metallicity-dependent stellar winds; an updated metallicity-dependent core-collapse supernova algorithm and a new supernovae fallback prescription; an updated electron-capture supernova and accretion-induced collapse and merger-induced collapse; high-mass remnant masses and natal kicks; BH natal spins, and other updates. For more details about our new stellar evolution library, we refer to \cite{Kamlah2022, Kamlah2022MNRAS, Banerjee2020}.

To carry out the $N$-body modelling together with the stellar evolution for selected GCs, we assumed the initial conditions listed below. 

\begin{itemize}

\item Because our TNG-TVP potential beyond eight billion years ago becomes quite unreliable, we started our GCs time-frame integration only after this time. Our choice of the initial integration time was mainly motivated by the external TNG potential characteristics. Formally, the TNG 100 potentials were calculated up to -11.6 Gyr, but the results of the first few billion years on the TNG halo evolution are very uncertain, especially for the selected present-day MW-like systems (see Fig. 2 in \hyperlink{I23a}{\color{blue}{Paper~II}}). These initial halo-disk and mass-size uncertainties can have a quite strong influence on the GC orbit backward integration beyond 8 Gyr from now. Based on this, we set our initial cluster positions and velocities 8 Gyr back in time. Initially, the GC stellar mass loss significantly dominates the dynamical tidal mass loss during the first very short $\approx100$~Myr time frame in the global mass evolution of the system. Due to this effect, the resulting differences between stellar mass loss starting at eight or even ten billion years lookback time is quite small in any case.

\item For the individual mass of the stars, we used the Kroupa initial mass function \citep[IMF;][]{Kroupa2001} with lower--upper mass limits equal to 0.08--100~$\rm M_{\odot}$. We also assumed a dynamical mass of each point as the mass of a particle ten times higher than the individual stellar mass of the represented star. Thus, one particle in the simulation represents a group of ten similar types of stars. Therefore, the stellar number counts were multiplied by 10 to be converted into real values (a similar approach has been applied to the simulation of the MW NSC dynamical evolution; see \citealt{Panamarev2019,Panamarev2018}, Section 2.2). This simplification allowed us to run a larger set of fitting models to derive the GCs initial mass and size parameters based on current observed values. To check the influence of this numerical particle reduction factor, we carried out two additional runs with a reduction factor of 20 and 5 for the cluster NGC~6642 (see the discussion in Section ~\ref{sec:in-gc}). 

\item Each cluster was initially set in a state of dynamical equilibrium with King model distribution function \citep{King1966}. These models are described by the two main initial parameters: the half-mass radius $r_{\rm hm}$, and the dimensionless central potential $W_{0}$ (see Table \ref{tab:init-param}). 

\item Each cluster had an individual metallicity according to Table~\ref{tab:phis-gc}, Column (6) \citep{Boyles2011}. Inside the code, to scale our stellar evolution models, we used the value of the classical solar metallicity $Z_{\rm m} \equiv Z_{\odot} = 0.02$ \citep{Grevesse1998}.

\item Initial coordinates and velocities at the 8 Gyr lookback time were taken from the orbital integration of the GC (see Table \ref{tab:init-param} and \hyperlink{I23a}{\color{blue}{Paper~II}}).

\item A Galactic potential {\tt 411324} TNG-TVP with the additional NSC Plummer-type potential \citep{Plummer1911} with a mass of $4\times10^{7}\rm\;M_\odot$ and a scale radius 4~pc were applied \citep{Schodel2014a, Fritz2016}. We did not use any special particle for the central SMBH potential.

\item During the star cluster initialisation, we did not add any primordial binary stellar component to the system. To avoid the very close dynamical encounter between individual stars during the evolution of the star cluster, we used a small enough softening parameter $\epsilon=0.1$~pc in the code. 

\end{itemize}

\subsection{Integration}\label{subsec:integr}

Most of the simulations were performed on the computer cluster at Fesenkov Astrophysical Institute, Almaty, Kazakhstan, under the TRIUNE project and on the LOTR cluster of the Main Astronomical Observatory of the National Academy of Sciences of Ukraine. Each run was calculated using a pair of Nvidia RTX 4080 or RTX 4090 GPU cards. On these cards, our simulations show a $\approx$21.2 Tflops sustained real performance on average. More than ten runs were also carried out on the JUWELS (Jülich Supercomputing Centre) supercomputer cluster, implemented under the madnuc project. We carried out numerical simulations to find a well-fitted individual model for each GC by varying the main cluster parameters, such as the mass $M$, the half-mass radius $r_{\rm hm}$, and the King concentration parameter $W_0$. 

We varied these initial parameters to discover the optimal conditions that following an eight-billion-year dynamical evolution would allow us to obtain the present-day observed cluster mass and half-mass radii with a deviation of $\pm$5\% in general. The fitting of the half-mass radius was usually a more time-consuming part of fitting the numerical runs. The value of the initial cluster mass was usually reproduced in a simpler and quicker way. 

The total integration time of our six selected clusters in parallel mode took one year and two months. The best-fitting initial models with individual $M$, $r_{\rm hm}$, and $W_0$ parameters for each GC are given in Table~\ref{tab:init-param}.

\begin{table*}[tbp]
\setlength{\tabcolsep}{4pt}
\centering
\caption{Initial kinematics and physical characteristics at a lookback time of eight billion years for the GCs.}
\label{tab:init-param}
\begin{tabular}{ccccccccccccc}
\hline
\hline 
GC & $X$ & $Y$ & $Z$ & $V_x$ & $V_y$ & $V_z$ & $E/m$ & $L_{\rm tot}/M$ & $M$ & $N$ & $r_{\rm hm}$ & $W_0$ \\
 & pc & pc & pc & km~s$^{-1}$ & km~s$^{-1}$ & km~s$^{-1}$ &  $10^{4}$ km$^{2}$~s$^{-2}$ & $10^{2}$~kpc~km~s$^{-1}$ & $10^{6}\rm\;M_{\odot}$ & & pc & \\
(1) & (2) & (3) & (4) & (5) & (6) & (7) & (8) & (9) & (10) & (11) & (12) & (13) \\
\hline
\hline
NGC~6401    & -1493 & -3213 & 398 & 5 & -3 & -57 & -18.8 & 2.03 & 1.2 & 2 249640 & 6.5 & 9 \\
Palomar 6   & 1499 & 465 & -1655 & 140 & 51 & -81 & -18.8 & 1.14 & 1.0 & 1 751 970 & 3.5 & 9 \\
NGC~6681    & 1392 & -4638 & 4243 & -0.7 & -1 & 61 & -16.6 & 2.9 & 1.3 & 2 265 380 & 3.0 & 8 \\
NGC~6642    &  983 & -897 & 545 & -47 & 54 & -117 & -21.2 & 1.1 & 1.5 & 2 613 856 & 4.0 & 9 \\
NGC~6981    & 1774 & 4001 & -4943 & 70 & 159 & -270 & -11.5 & 3.7 & 1.0 & 1 742 560 & 7.0 & 9 \\
HP 1        & -580 & 242 & 1859 & 71 & -36 & 128 & -19.5 & 2.3 & 1.3 & 2 265 340 & 6.0 & 8  \\
\hline 
\end{tabular}
\tablefoot{
Column (1) -- the GCs names;  
columns (2) -- (4) -- initial position in Cartesian galactocentric frame;
columns (5) -- (7) -- initial velocities in Cartesian galactocentric frame;
column (8) -- total specific energy; 
column (9) -- total specific angular momentum;
column (10) -- initial mass;
column (11) -- initial number of stars;
column (12) -- initial half-mass radius;
column (13) -- initial concentration of King profile.
}
\vspace{6pt}
\end{table*}

\section{Inner and global evolution of the globular clusters}\label{sec:inn-glob-evol}

\subsection{Inner structure of the globular clusters}\label{sec:in-gc}

\begin{figure}[ht]
\centering
\includegraphics[width=0.99\linewidth]{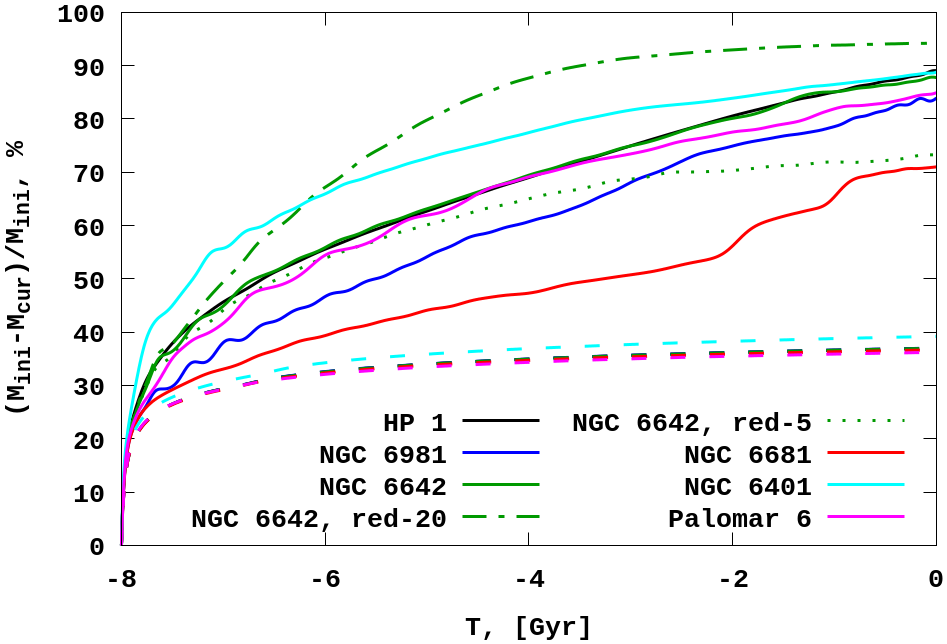}
\caption{Evolution of the mass loss in the GCs in percent due to mass loss and orbital type. The solid lines represent the full mass loss of the clusters with a reduction factor of 10. The dotted and dot-dashed green lines represent the NGC~6642 models with reduction factors of 5 and 20. The dashed lines represent the mass loss that is only due to the stellar evolution.}
\label{fig:mass-loss}
\end{figure}

We followed the detailed dynamical and stellar mass loss of our set of GCs. First of all, we compare the mass-loss evolution of our clusters in Fig.~\ref{fig:mass-loss}. The total mass loss is quite similar for all GCs. Typically, 70-90\% of the initial cluster mass is lost during the entire time evolution of 8~Gyr. 

Clusters with a similar Galactic angular momentum and bounding energy values also typically show a similar mass-loss behaviour (see Table~\ref{tab:init-param} and e.g. NGC~6642 and HP~1). We show the total mass loss of the clusters: the dynamical mass loss due to the external time variable potential, plus the stellar mass loss due to the stellar evolution of the stellar GC population alone in Fig. \ref{fig:mass-loss} (solid lines). The stellar evolution mass loss appears to be largely independent of metallicity, despite initial differences among the GCs. Within the first $\approx 100$~Myr, the GCs lose approximately 25\% of their initial masses. This early mass-loss phase is predominantly due to stellar evolution. The subsequent mass loss is primarily attributed to tidal interactions with the external potential. NGC~6681 effectively has the lowest total mass-loss rate of our selected clusters. As a consequence, it still has $\approx$70\% of the initial stellar mass of the cluster today.

\begin{table}[htbp]
\caption{Masses inside $r_{\rm hm}$ and $r_{\rm tid}$ for today, according to the numerical modelling.}
\centering
\begin{tabular}{cccccc}
\hline
\hline
GC & $r_{\rm hm}$ & $r_{\rm tid}$ & $M (r_{\rm hm})$ & $M (r_{\rm tid})$  \\
& pc & pc & {$10^{4}\rm\;M_{\odot}$} & {$10^{4}\rm\;M_{\odot}$} \\
\hline
\hline
NGC~6401  & 4.3 & 28  & 7.1 & 14.2 \\
Palomar~6 & 2.7 & 70  & 7.6 & 15.2 \\
NGC~6681  & 4.5 & 37  & 18.5 & 37.0 \\
NGC~6642  & 1.5 & 11  &  8.2 & 16.6 \\
NGC~6981  & 8.9 & 215 & 10.4 & 15.9 \\
HP~1      & 4.0 & 37  &  7.0 & 14.2 \\
\hline
\end{tabular}
\label{tab:rhm-mass}
\end{table} 

As we discussed in Section~\ref{subsec:init-cond} in item two, we used a scaling factor of 10 in the number of particles for our simulations (similar as in \citealt{Panamarev2018,Panamarev2019}, Section~2.2). To determine the influence of this reduction factor in a particle number on the results of our numerical modelling, we additionally ran two more runs for NGC 6642, with reduction factors of 5 and 20. Fig.~\ref{fig:mass-loss} shows that all three models with significantly different reduction factors have quite similar mass-loss properties. The larger reduction factor (20) causes some large deviation for the standard model (with a reduction factor of 10), especially between -6 and -2~Gyr. At the end of the simulation (i.e. today), however, the final masses of these models differ by only a few percent. 

The differences between the mass loss for runs with reduction factors of 10 and 5 are remarkably small up to $\approx$-4~Gyr, except for the difference of a factor of two in the particle numbers. Even at the end of the simulation, the general mass loss for these above runs differs only at a scale of 17\%. For the purpose of studying the long-term general mass loss of GCs in the close vicinity of our GalC, this numerical deviation is quite acceptable when we take the huge (factor of $\approx$100) running speed of the reduced compared to the not reduced models into account. 

\begin{figure}[ht]
\centering
\includegraphics[width=0.99\linewidth]{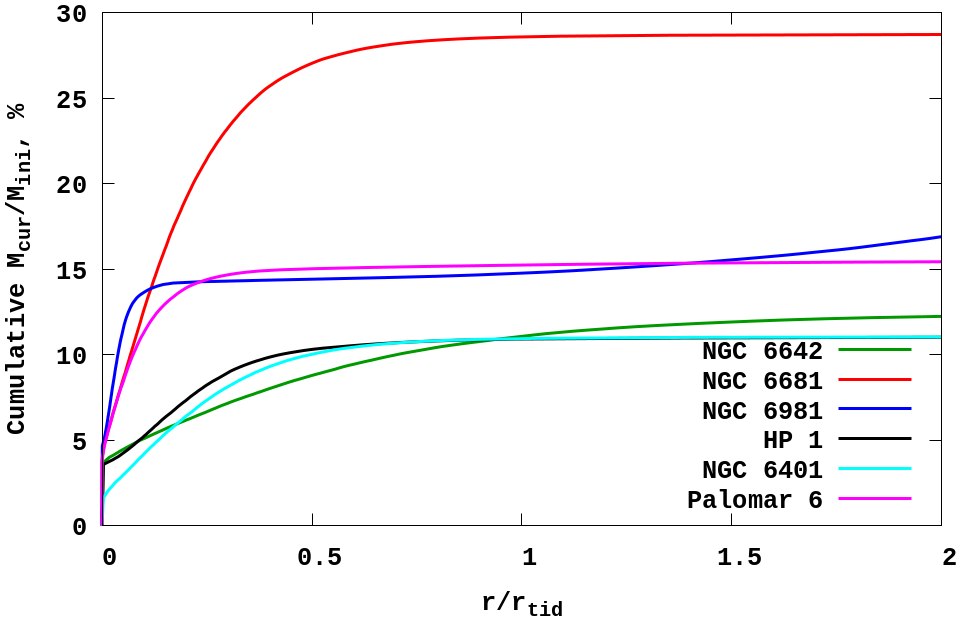}
\caption{Internal cumulative mass distribution of the GCs as a function of the distance in units of the tidal cluster radius at present according to our numerical simulations.}
\label{fig:mass-distr}
\end{figure}

In Fig.~\ref{fig:mass-distr} we present the internal mass distribution of the GCs as observed today. The current tidal radii for our set of GCs according to the numerical modelling is presented in Table~\ref{tab:rhm-mass}. The tidal cluster radius (or Jacobi or King radius) was calculated based on the numerical iteration of the $M_{\rm tid}$ and $r_{\rm tid}$ values in the equation
\be
r_{\rm tid} = \left[ \frac{G \cdot M_{\rm tid}}{4\Omega^2-\kappa^2} \right]^{1/3}, 
\ee
where $G$, $M_{\rm tid}$, $\Omega$, and $\kappa$ are the gravitational constant, the tidal cluster mass, and the circular and the epicyclic frequencies of a near-circular orbit in the Galaxy at the current GC position, respectively. For more details, we refer to \cite{King1962} and \cite{Ernst2011}. 

The wide range of the current cluster $r_{\rm tid}$ is conditioned by the large spread of GC distances from the GalC. Comparing the numerical model half-mass radii and tidal radii, we can conclude that they are in the range of the observed values (see Table~\ref{tab:phis-gc} and \cite{Baumgardt2021}). 

The mass distribution of the clusters is quite compact. Some of the clusters, NGC~6642, NGC~6401 and HP 1, have a very compact ($r<0.1r_{\rm tid}$) almost isothermal core ($\rho\sim1/r^{2}$). Almost all clusters have a sharp edge of the stellar systems around 
$r\approx r_{\rm tid}$, but the mass distribution of NGC 6981, for example, extends slightly above $r_{\rm tid}$ because the unbound tidal tails of the system are extended.     

\subsection{Dynamical evolution of the high-mass remnants}\label{sec:rem}

We paid special attention to the compact stellar remnants within this central region. Using the detailed stellar evolution modelling in our $N$-body code, we monitored the internal kinematics and long-term orbital evolution of the stellar remnants, mainly neutron stars (NS) and black holes (BH) \citep{Kamlah2022}. The dynamically most significant events during the lifetimes of high-mass stars are their final supernova explosions. These highly energetic events typically mean that the newly formed stellar remnant receives a randomly oriented, fallback-scaled natal kick \citep{Banerjee2020}.

\begin{figure}[ht]
\centering
\includegraphics[width=0.99\linewidth]{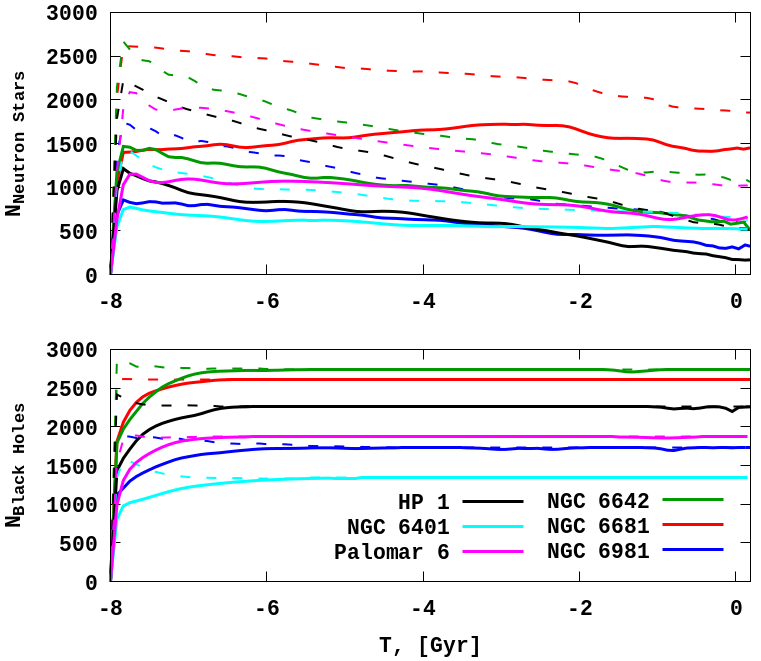}
\caption{Time evolution of the compact stellar remnants in the central part of the GCs up to $r_{\rm hm}$ (solid lines) and inside $r_{\rm tid}$ (dashed lines).  Neutron stars are depicted in the \textit{upper panel}, and black holes are shown in the \textit{bottom panel}.}
\label{fig:loc-bh}
\end{figure}

The time evolution of the total number of BHs within a GC is presented in the bottom panel of Fig.~\ref{fig:loc-bh}. Initially, the number of bound BHs (inside $r_{\rm tid}$ with the bound status) becomes saturated during the first billion years of the GC evolution. During the first $\approx$100~Myr, the total number of BHs increases rapidly. In the subsequent few hundred million years, the number of bound BHs steadily decreases by approximately 10\%. The global number of BHs (both bound and unbound) for each GC is presented in Table~\ref{tab:ns-bh}. The proportion of bound BHs is approximately 60-65\%. The number of BHs in the central region of the clusters ($r < r_{\rm hm}$) follows a pattern similar to that of the bound BHs, except during the first billion years of evolution. However, after this period, all the bound BHs are observed to form within the half-mass radius of the so-called BH su system in the central cluster core \citep{BH2013a, BH2013b, Mor2015, Wang2016}. It is evident that the fraction of escaped BHs is a direct consequence of the kick algorithm (including fallback) applied in our numerical code \citep[see Section 4.2.3]{Kamlah2022}.

In Fig.~\ref{fig:loc-bh} (\textit{upper panel}) we also present the time evolution of the total number of NSs within GCs, both within $r < r_{\rm tid}$ and inside the half-mass radius. The rapid increase (within less than 100~Myr) in the number of NSs inside the clusters is a noticeable trend. It is followed by a steady decline. The global count of NSs (both bound and unbound) for each GC is detailed in Table~\ref{tab:ns-bh}. The proportion of bound NSs is approximately 5-10\% as of today. This limited number of bound NSs can be attributed to the kick algorithm that is implemented in all NS formation scenarios \citep[see Section 4.2.3]{Kamlah2022}. The discrepancy between the numbers of NSs within the tidal radius and within the half-mass radius contrasts with the distribution of black holes at similar distances. Specifically, only about 50--70\% of the initially formed NSs are found within the half-mass radius. This indicates that a substantial fraction of NSs are loosely bound to the cluster, unlike the BH subsystem.

In Table~\ref{tab:ns-bh} we present the total number of NSs and BHs that formed over the 8~Gyr dynamical evolution of our GCs. The total number of NSs constitutes approximately 0.7\% of the initial star count, while the total number of BHs is about 0.2\%. Despite the small number fraction of these remnants, they play a crucial role in the internal dynamical evolution of GCs \citep{Wang2016, Kamlah2022MNRAS}. These high-mass remnants, in particular, the BHs, undergo mass segregation. They migrate to the central region to form a relatively stable BH subsystem \citep{Rizzuto2021, Arca2023}.

\begin{table}[htbp]
\caption{Total number of NSs, BHs, and BH systems that formed in GCs.}
\centering
\begin{tabular}{cccccc}
\hline
\hline
GC & NS & BH & $N_{\rm sub-sys, BH}$ & $M_{\rm sub-sys, BH}, \rm M_{\rm \odot}$  \\
\hline
\hline
NGC~6401  & 12 480 & 3 570 & 1340 & $1.99 \times 10{^4}$ \\
Palomar~6 & 13 540 & 3 020 & 1870 & $3.99 \times 10{^4}$ \\
NGC~6681  & 15 910 & 3 970 & 2610 & $5.61 \times 10{^4}$ \\
NGC~6642  & 17 730 & 4 440 & 2740 & $5.82 \times 10{^4}$ \\
NGC~6981  & 11 710 & 2 990 & 1720 & $3.66 \times 10{^4}$ \\
HP~1      & 15 140 & 3 890 & 2260 & $4.76 \times 10{^4}$ \\
\hline
\end{tabular}
\label{tab:ns-bh}
\end{table} 

\begin{figure}[htbp!]
\centering
\includegraphics[width=0.99\linewidth]{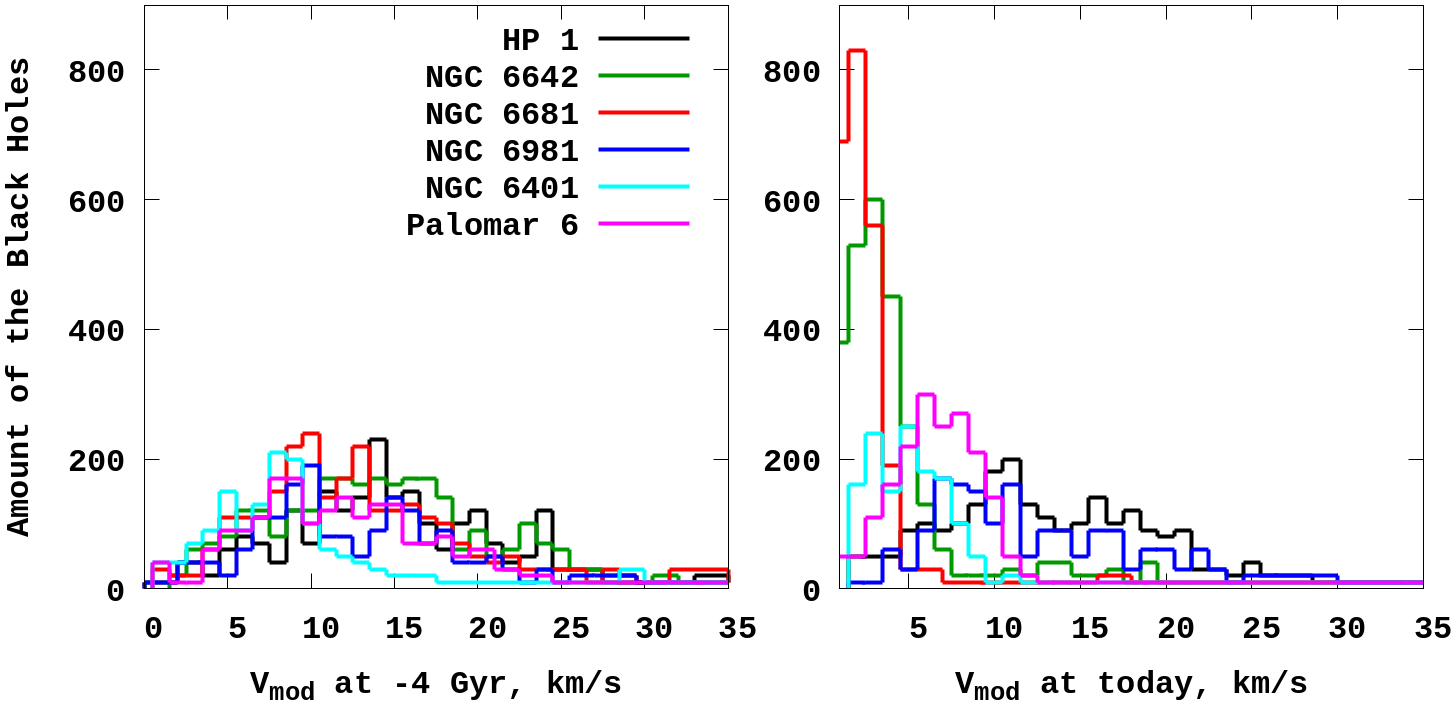}
\includegraphics[width=0.99\linewidth]{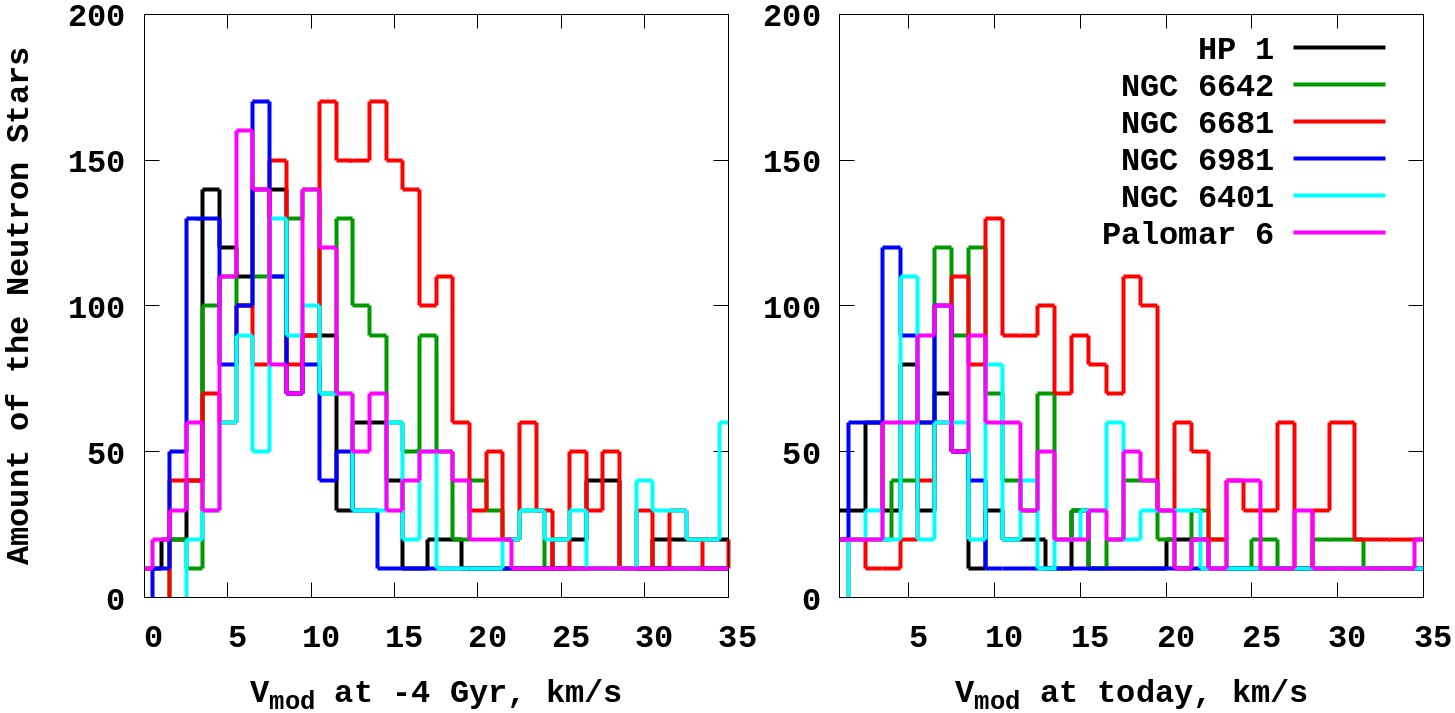}
\caption{Velocity modulus distribution of BHs and NSs relative to the GC centre. We present the distribution at two moments of time: at four Gyr lookback time in the \textit{left panel}, and for today in the \textit{right panel}.}
\label{fig:dv-bh}
\end{figure}

The clear manifestation of dynamical subsystems of the compact objects for two selected moments of time as a function of velocity modulus in a cluster centre frame is shown for all our selected GCs in Fig.~\ref{fig:dv-bh}. As expected from our fallback prescription \citep{Banerjee2020, Kamlah2022}, the NSs distributions are much wider and generally follow a Maxwellian type of distribution with a long tail. For the current four clusters (NGC~6681, NGC~6642, NGC~6401, and  Palomar~6), the concentration of BHs in the cluster centre velocity frame is clear. This accumulation process in these GCs already starts after about five billion years of the cluster evolution. For these objects, a significant fraction of BHs forms centrally peaked BH subsystems \citep{Wang2016, Rizzuto2021, Arca2023}. For the NS remnants, according to our fallback prescription, the natal kick is always strong \citep{Banerjee2020, Kamlah2022}, so that only a relatively small fraction of NS can stay inside the cluster. 

\begin{figure}[htbp!]
\centering
\includegraphics[width=0.99\linewidth]{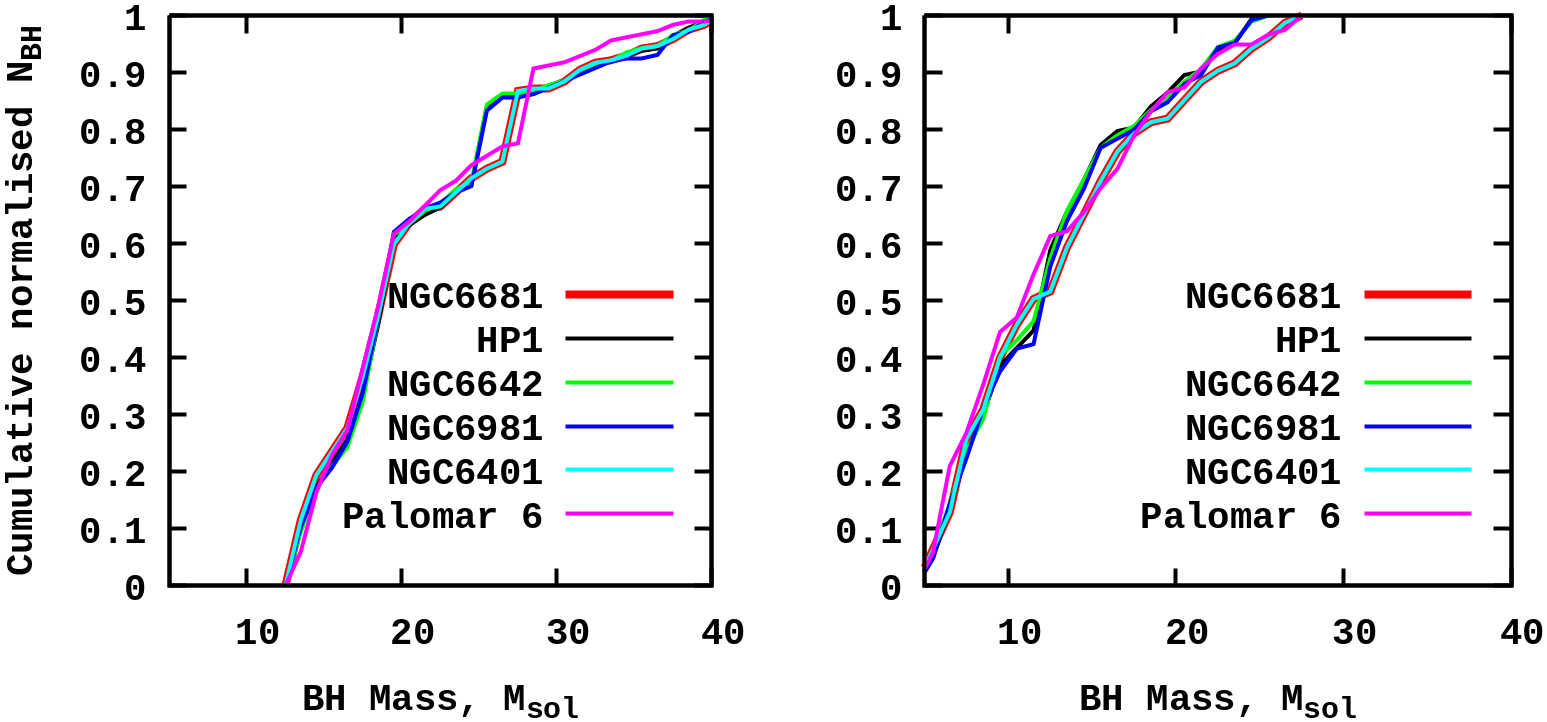}
\caption{Statistics of the natal-kick dependence from the BH remnant final masses.  $V_{\rm kick}=0$ is shown in the left panel, and $V_{\rm kick}>0$ is shown in the right panel.}
\label{fig:kick}
\end{figure}

In Fig.~\ref{fig:kick}, we present the difference in the final mass distributions of BHs as a function of the presence or absence of the natal kick at the moment of BH formation within all six clusters. The total numbers of the BH remnants in different clusters can be found in Table~\ref{tab:ns-bh}. As we expect from our kick prescription, the high-mass BHs ($\approx15-40\rm\;M_{\odot}$) generally receive almost no kick, whereas the lower-mass BHs ($\approx5-25\rm\;M_{\odot}$) are subjected to a non-zero kick.

\subsection{Global evolution of the globular clusters}\label{sec:glob}

\begin{figure*}[htbp!]
\centering
\includegraphics[width=0.99\linewidth]{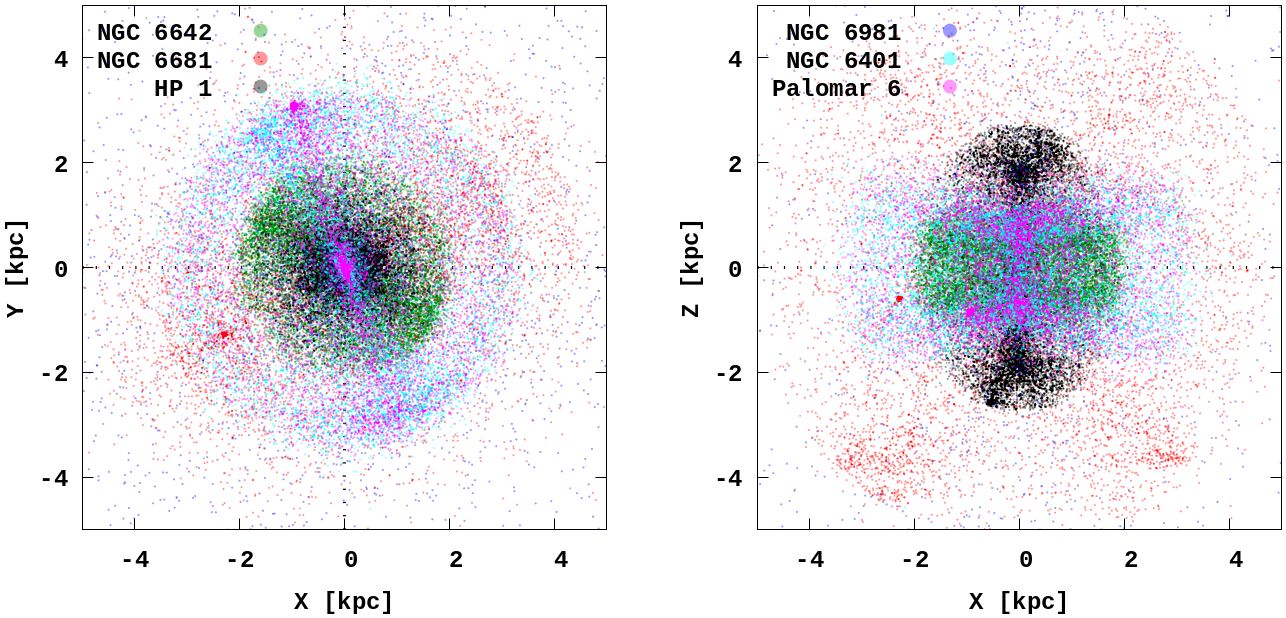}
\caption{Projected positions of the stars for six GCs in the central region of the Galaxy at the present time.}
\label{fig:pos-gc}
\end{figure*}

In Fig.~\ref{fig:pos-gc} we present the positions of all (bound and unbound with a GC) the stars from our six clusters in the central region of the Galaxy at the present day, based on our numerical simulations. Two GCs with a radial orbit (NGC~6981 and NGC~6681) have a smeared density structure in the Galaxy. These objects are presented as blue and red dots. The stars from HP~1, which have an irregular orbit, show a stellar density core in a central box $\pm1$~kpc in ($X,Y$) projection, and they show concentrated density areas in the \textit{Z} direction at a distance up to 2.5~kpc in both directions along the $Z$-axis.

On the other hand, NGC~6642, which has a tube-shaped orbit, has a different distribution of the stellar density (green dots). A concentrated tube-like density region is located at $\pm2$~kpc in ($X,Y$) and in $\pm0.8$~kpc in the $Z$-direction. At the same time, NGC~6401 and Palomar 6 (cyan and magenta dots), which similarly have a tube orbit, lack an increased stellar density in the central region of about one kiloparsec.

\begin{figure*}[htbp!]
\centering
\includegraphics[width=0.99\linewidth]{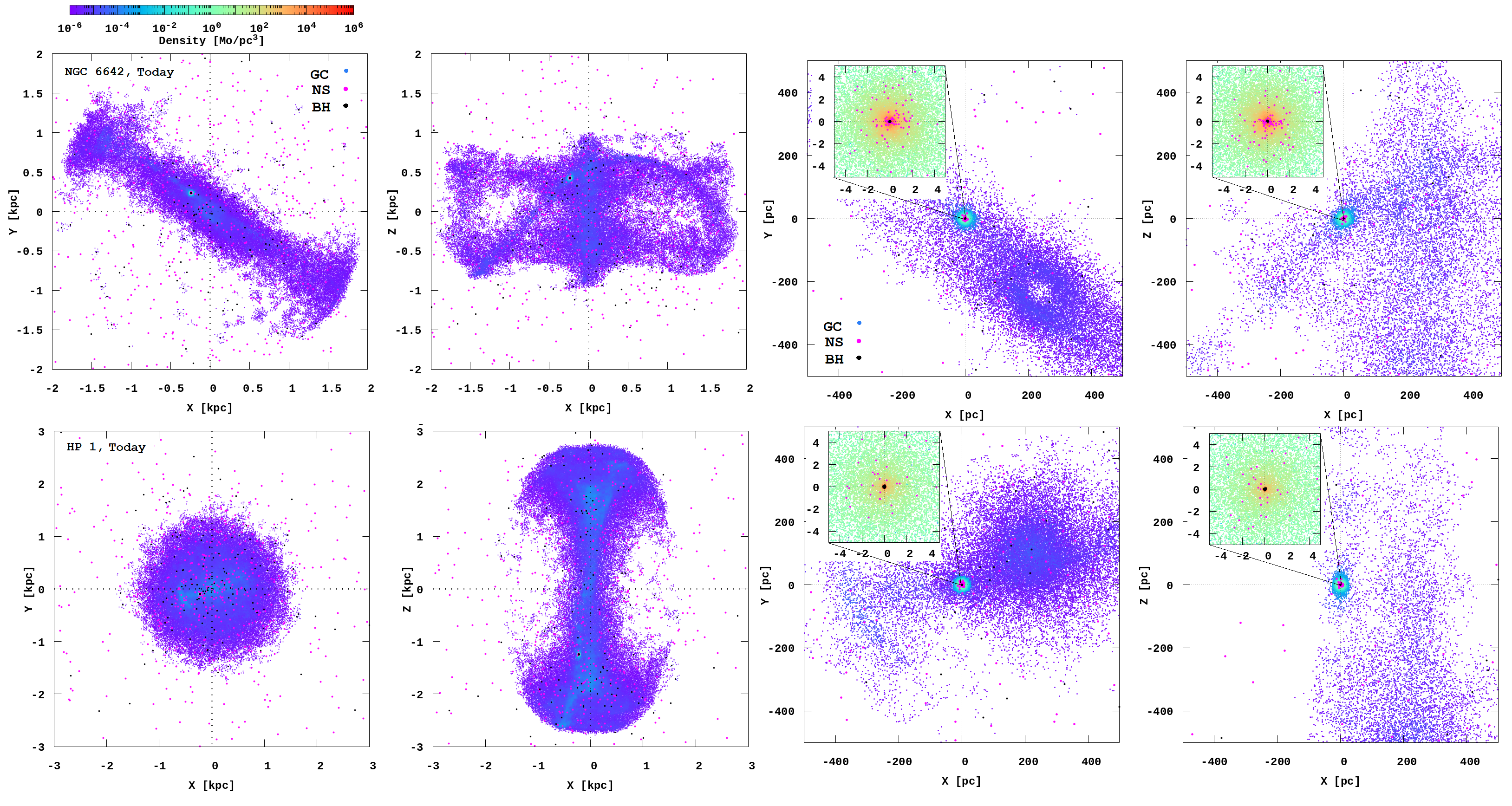}
\caption{Clusters NGC~6642 and HP~1. We show the current density distributions in the{\tt 411321} TNG-TVP external potential. The orbital global evolution is present in two projections ($X$,$Y$) and ($X$,$Z$) in the \textit{left panels}. The central GC part in the cluster local frame with the BHs (black dots), NSs (magenta dots), and with a detailed zoom of the central area (with box size $10$~pc) is shown in the right \textit{two panels}.}
\label{fig:gc-today}
\end{figure*}

In Fig.~\ref{fig:gc-today} we present the large-scale and zoomed view of the current density distributions for two of our selected clusters: NGC 6642 and HP 1. In the left panel of the figure, we show the global Galaxy-centred view of the particle density distributions. In the right panel, we present the cluster-centred local distributions. Within the zoomed frames on the right sides, we show the GC central $\pm$5 pc box with the particle distributions. In all the plots, the BHs are especially highlighted as black dots and the NSs are shown as magenta dots. 

The global orbital evolution of all the stars from our $N$-body modelling can be found in Zenodo service\footnote{GCs global stellar density distributions: \\~\url{https://zenodo.org/records/12107255}}. In general, the orbits of our clusters are quite complex, but due to our selection criteria, they all have a close passage near the GalC. The global behaviour during the whole time of integration is presented in the left panels. For all objects, the over-density of the striped stars in the GalC region is significant. These stars can be potentially used as building blocks for the Galactic NSC. A significant feature in the global tidal tails are shell-like structures. These structures are clearly connected with the bounding energy and orbital apocentre turning points of the clusters. 

The illustrations show that these structures last long (except for HP~1). These shells are hard to observe because their densities are quite low: $\approx$ 0.01 M$_\odot$/pc$^3$. For comparison, the background mass density in the solar neighbourhood is $\approx$ 0.059 M$_\odot$/pc$^3$ (see \cite{erik2009milky}). These structures are therefore well beyond the average stellar mass density in the Galaxy. These shell-like structures can probably only detected using the full 6D (spatial and kinematic) characteristics of the MW stellar content.

The detailed view of clusters structure including the high stellar mass remnants (NSs and BHs) is presented with separate dots in the right panels. For each detailed view of a GC, we also prepared a central 10~pc zoomed box image of the densities. For NGC~6642 and HP~1, a clear signature of the density clumps in the large-scale tidal tails in the range of $\approx200$~pc from the cluster centres is visible. The existence of these density clumps in the tidal cluster tails was demonstrated in our earlier work \cite{Just2009}.

\begin{figure*}[htbp!]
\centering
\includegraphics[width=0.99\linewidth]{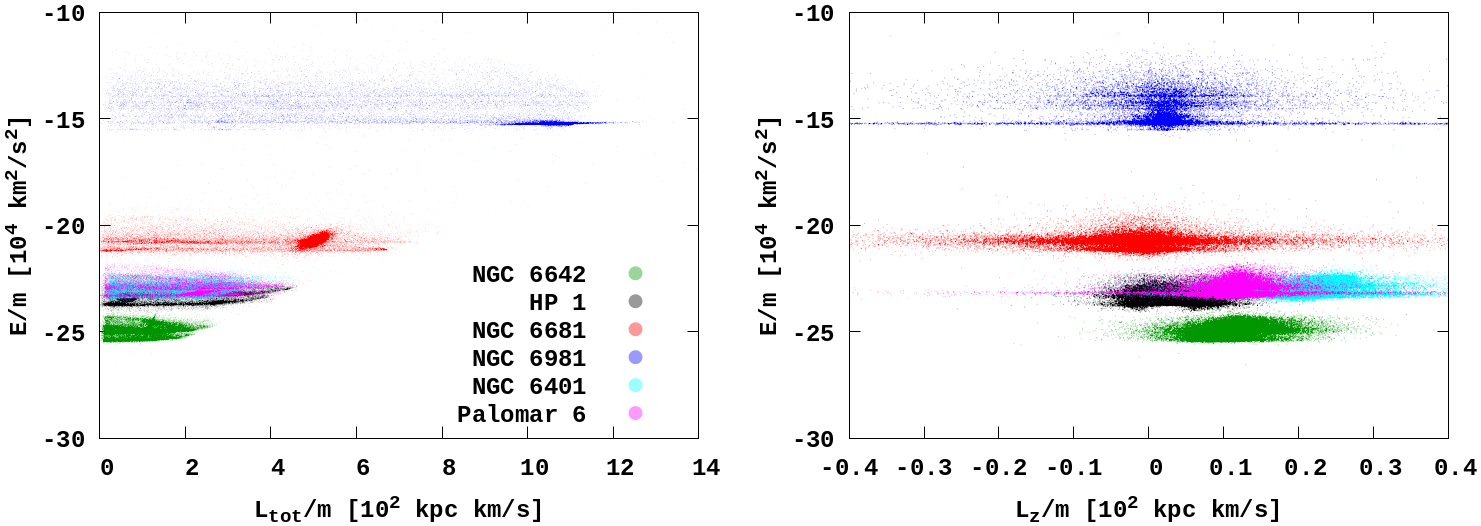}
\caption{Total specific energy ($E/m$) vs. total specific angular momentum ($L_{\rm tot}/m$) and total specific energy vs. $z$th component of the specific angular momentum ($L_{\rm z}/m$) for the current GCs.}
\label{fig:e-ltot}
\end{figure*}

In addition, we also analysed the current phase-space distribution of all the stars from all the clusters. The corresponding data are shown in Fig.~\ref{fig:e-ltot}. We present the specific total stellar energy ($E/m$) and specific angular momentum data ($L_{\rm tot}/m$ and $L_{\rm z}/m$) with the same colour-coding as in the previous figures. From these figures, for the purpose of our investigation as a search for the potential source of stars from GC to the NSC, we expect that clusters NGC~6642, HP~1,  NGC~6401 and Palomar~6 may have been a possible source of the stellar population of the NSC in the past (see the details in Section~\ref{sec:int-nsc}).

From an observational perspective, the high-mass remnants of GCs (e.g. NSs and BHs) are highly energetic objects that are distributed throughout the Galaxy. To analyse these remnants, we specifically studied these objects in the Galactic field.
The standard model for the formation of millisecond pulsars indicates that they originate from neutron stars in binaries, typically paired with a red giant or a main-sequence star. Although our code, \texttt{PhiGPU}, does not include prescriptions for binary stellar evolution, we can still provide insights into the potential presence and kinematic properties of millisecond pulsars in the central region of the MW that originated from the GCs we studied in this paper.

\begin{figure}[htbp!]
\centering
\includegraphics[width=0.95\linewidth]{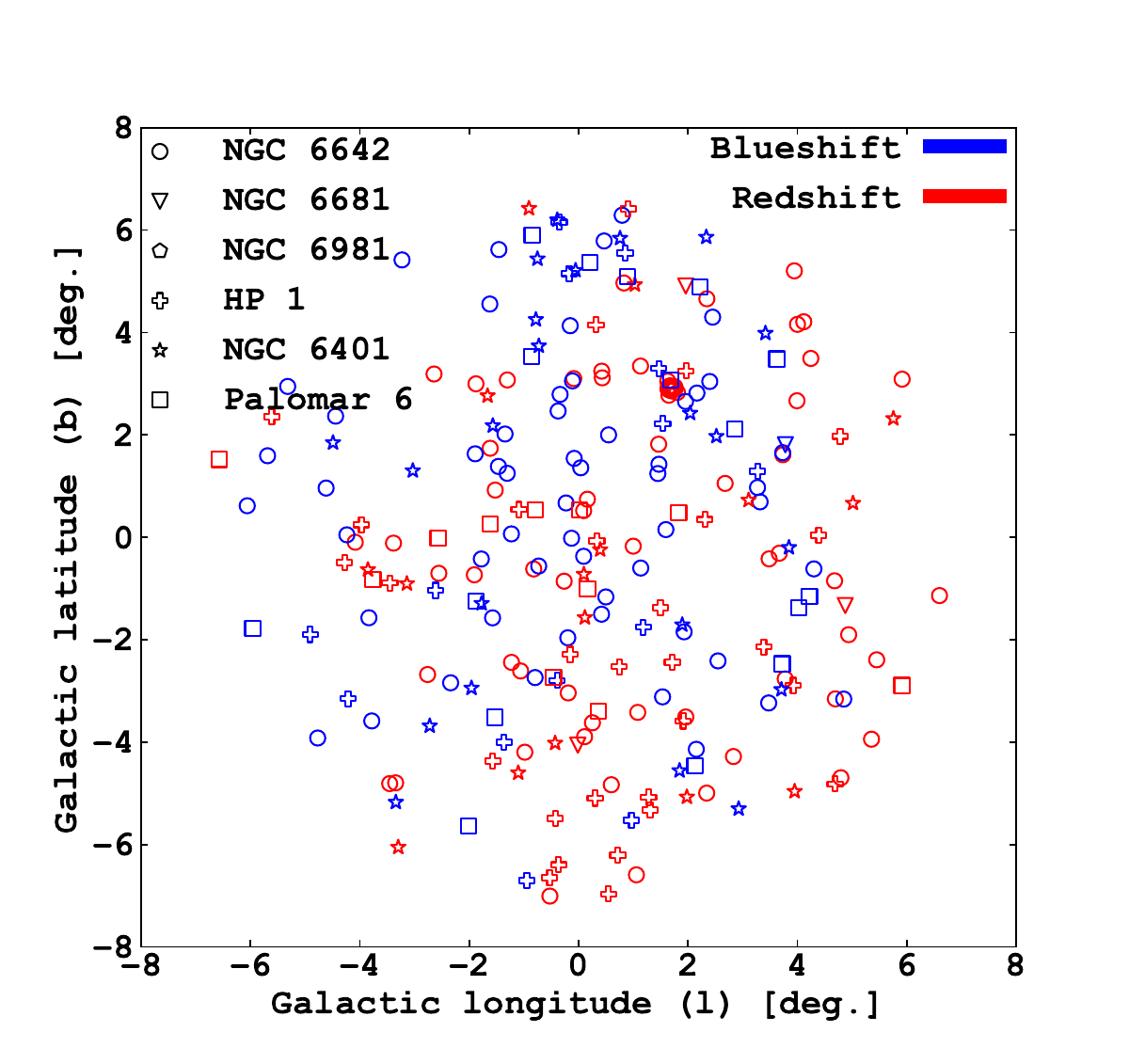}
\caption{Neutron star distributions within a sphere of 1~kpc from the GalC projected in Galactic coordinates, with a box of $\pm8$-degrees for all current GCs.}
\label{fig:zoom-box-8}
\end{figure}

In Fig.~\ref{fig:zoom-box-8}, we present the NS distributions within a sphere of 1 kpc from the GalC for today, projected in Galactic coordinates with a box of $\pm 8$ degrees from all GCs. Blueshifted NSs are depicted in blue, and redshifted NSs are shown in red. In this area lie 2210 NSs from NGC 6642 (including 1110 NSs inside the cluster), 380 NSs from NGC 6401 \footnote{NGC 6401 orbital visualisation: \\~\url{https://zenodo.org/records/12107255}, NGC-6401.mp4}., 420 NSs from HP 1, 40 NSs from NGC 6681, and 270 NSs from Palomar 6. In Fig.~\ref{fig:zoom-box-8}, we present only every tenth point from each of these objects. No NSs from NGC 6981 lie in this area. The large number of NSs from NGC 6642 is understandable because the current centre of this cluster is occasionally inside this $\approx$ 1 kpc Galactic central region (the dense red clump in the upper right region of the plot).

In Appendix~\ref{app:NS-distr} we present the global NS distribution in the Galaxy. The colour-coding corresponds to the distance of individual NS from the Sun. The NS distribution on the sky is quite random in each direction, but is concentrated for all of our GC NSs in the range of $l$ between $-30^\circ$ and $+30^\circ$. In the case of long radial types of orbits (NGC~6981 and NGC~6681), $b$ ranges between $-90^\circ$ and $+90^\circ$. A significant number of NSs are located inside the sphere of $\approx15$~kpc around the Sun, but part of the NSs is distributed up to 100~kpc and even beyond.

\section{Probability of stellar accumulation onto the central Galactic NSC}\label{sec:int-nsc}

We discussed the dynamical processes of strong interaction between GC and GalC and their implication for the NSC formation and mass growth previously \cite{ArcaSedda2019}. In the present work, we significantly extend the reality of the simulations. First of all, we investigated the real GCs from our list of Galactic GCs. We significantly extended the internal structure of the GCs based on the set of fitting $N$-body models to the main parameters of present-day GCs. Additionally, we included a more realistic recent stellar evolution prescription \citep{Banerjee2020, Kamlah2022, Kamlah2022MNRAS}. The new stellar evolution routines provide us with significantly higher BH remnant masses and also with slightly different ratios of BH to NS numbers. 

In \cite{ArcaSedda2019} we studied the very close inner $\approx10$~pc GalC region and the dynamical dissolution of hypothetical GC on bound orbits around the central SMBH (see Figs.~17 and 18 herein). In comparison with our current study, we previously only analysed the full dissolution of hypothetically bound GC. In the current work, the mass accretion study is more detailed and also includes the high-mass remnant distribution around the forming NSC.    

During the simulations, we followed the individual star trajectories in the Galaxy and especially detected each event when the stars entered the inner GalC zone: $D < 100$~pc. We therefore separately recorded all cases when the stars entered or left the 100~pc sphere around the GalC. During the orbital evolution of the GCs, we neglected the dynamical friction forces acting on the stars. In previous studies, for example, \citealt[also see Eq. 1]{Just2011}, \citealt{Just2012}, some approximation equations were propsed to describe the effect of dynamical friction acting on the stars in the central Galactic region with different mass profiles. 

Based on these ideas, we introduce the relative velocity redaction factor, namely -- $V_{\rm red}$, during the analysis of the bound status of the stars of each individual interaction with the central NSC potential, based on the snapshot data after the dynamical simulations were already completed.  

Because we neglected the possible dynamical friction effects on the GC and individual stellar motions in our simulations, our NSC passing velocities are at the very upper limits. To roughly estimate the possible velocity loss effects, we mimicked the complex friction with the simple velocity redaction for the stars during the close interaction with the NSC (in first-order approximation),
\begin{equation}
E_{\rm bound}^{\ast} = -(M_{\rm NSC} + m_{\ast}) / R_{\ast} + V_{\rm red} \cdot 0.5 \cdot V_{\ast}^2 , 
\label{eq:e_bound}
\end{equation}
where E$_{\rm bound}^{\ast}$ is the star-specific gravitational energy due to the NSC interaction. $M_{\rm NSC}$ is the NSC mass, $R_{\ast}$ is the separation between the NSC and the stars, and $V_{\ast}$  is the relative velocity between them. We introduced this additional V$_{\rm red}$ factor to take the relative velocity reduction due to the global GalC dynamical friction effects in the central dense environment into account. 

\begin{figure*}[htbp!]
\centering
\includegraphics[width=0.45\linewidth]{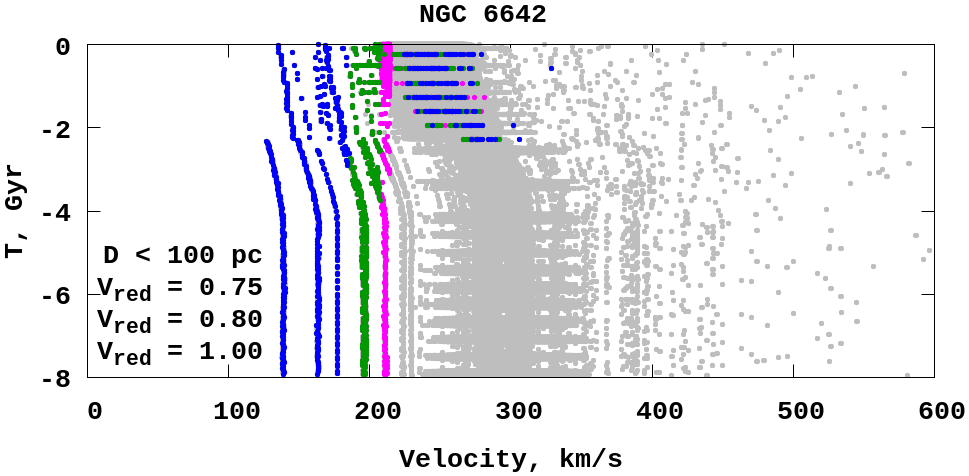}
\includegraphics[width=0.45\linewidth]{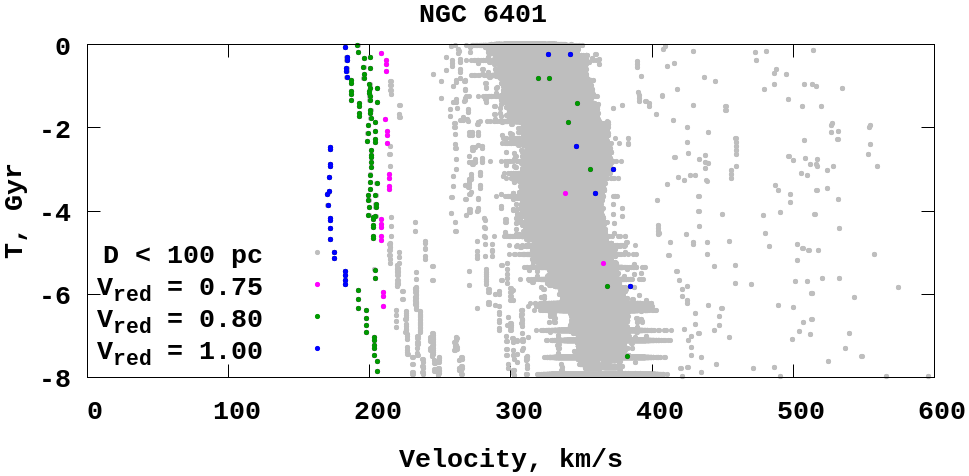}
\includegraphics[width=0.45\linewidth]{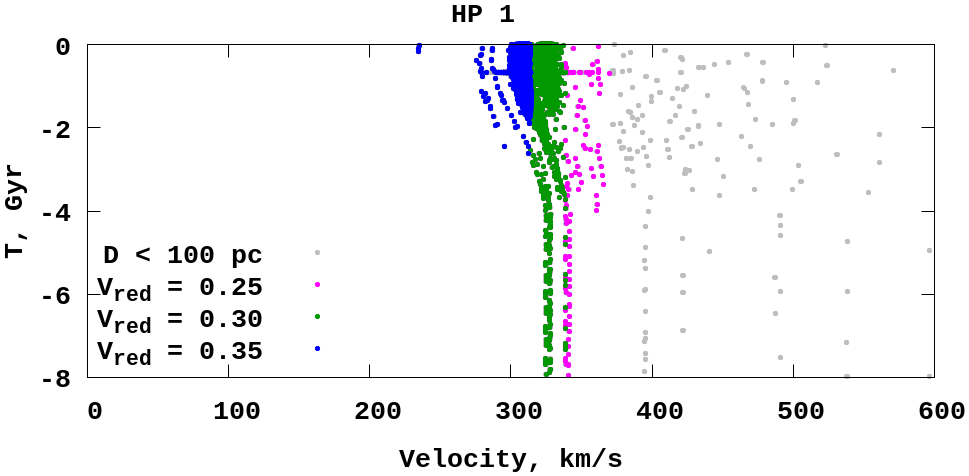}
\includegraphics[width=0.45\linewidth]{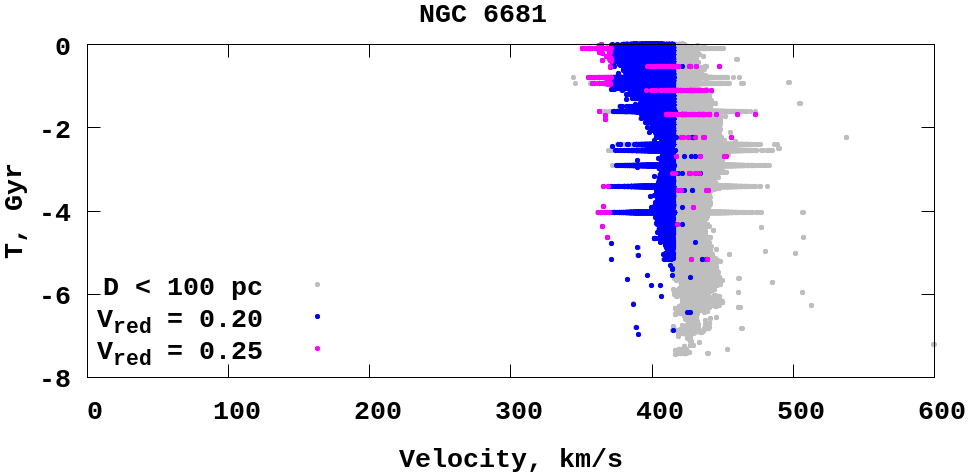}
\caption{Velocity distribution during the full integration time for particles that closely pass near the NSC at less than 100~pc (grey dots). Particles with a bound status ($E_{\rm bound}^{\ast} < 0$) with different velocity reduction factor values $V_{\rm red}$ are represented by coloured dots. NGC~6642 and NGC~6401 are shown in the \textit{upper panel}, and HP~1 and NGC~6681 are shown in the \textit{bottom panel}.}
\label{fig:red-vel}
\end{figure*}

In Fig.~\ref{fig:red-vel} we show the velocity map during the full integration time for particles that closely passed near the NSC ($D < 100$~pc). These particles are presented as grey dots. The relative to GalC velocity of the GC cores is about $300\pm70$~km~s$^{-1}$. We present the data for the four GCs NGC~6642, NGC~6401, HP~1, and NGC~6681.

For a possible detection and analysis of GC bound particles with NSC, we applied several velocity factor reductions for NGC~6642. This GC has the lowest crossing velocity in our GCs sample, $290\pm50$~km~s$^{-1}$. It is important to note that even without the velocity reduction factor ($V_{\rm red}$ = 1.0), we already have a set of bound particles with NSC (black dots). For $V_{\rm red}$ = 0.80, 0.75, the bound particle numbers increase by two to four times (dark green and magenta). With the smaller additional velocity factor, up to $V_{\rm red}$ = 0.60, 0.50, almost all particles from NGC~6642 become bound with the NSC (see Table~\ref{tab:ptcl}). We directly conclude that only for the NGC~6642 do we have some NSC bound particles from the GC inside the $\lesssim100$~pc with $V_{\rm red} = 1.00$. 

For NGC~6681 and HP~1, the core velocities are about $410\pm35$~km~s$^{-1}$ and $325\pm10$~km~s$^{-1}$ during the whole dynamical evolution. We present this in  Fig.~\ref{fig:red-vel} with the colour-coding. For these objects, the bound particles with NSC are practically not observed for $V_{\rm red}$ from 1.0 to 0.50. Only for the velocity reduction factor within the range of $V_{\rm red}$ = 0.35 -- 0.20 do increasingly more particles become bound. Only in this case do we reach some significant fraction of bound particles for these objects (Table~\ref{tab:ptcl}). 

To summarise Fig.~\ref{fig:red-vel}, we present the values in Table~\ref{tab:ptcl}. We list the number count of the close NSC passing particles for the three typical orbit shapes. In our sample, NGC~6642 and NGC~6401 represent the tube orbits, NGC~6681 represents the radial orbit, and HP~1 represents the irregular orbit type. All of these GCs dynamically belong to the Galactic bulge subsystems. We show the total number of stars with a bound status with the NSC in the moment of the closest passing during the entire eight billion years of integration. Because some of the stars enter the sphere close to NSC 100~pc multiple times, we also show the number of unique bound stars during the integration (ignoring the multiple passages). These values are marked in Table~\ref{tab:ptcl} as ``Tot.'' (total) and ``Unq.'' (unique) numbers, with the percent from the initial number of all stars in GCs.

\begin{table*}[htbp]
\caption{Total number of the bound particles with the NSC and their masses with different velocity redaction during eight billion~years of evolution in units of $10^3$.}
\centering
\begin{tabular}{c|ccc|ccc|ccc|cccc}
\hline
\hline
Vel$_{\rm red}$ & \multicolumn{3}{c}{NGC~6642} & \multicolumn{3}{c}{NGC~6401} & \multicolumn{3}{c}{NGC~6681} &  \multicolumn{3}{c}{HP~1} \\
    & Tot. & Unq. / in \% & M$_{\rm \odot}$ & Tot. & Unq. / in \% & M$_{\rm \odot}$ & Tot. & Unq. / in \% & M$_{\rm \odot}$ & Tot. & Unq. / in \%  & M$_{\rm \odot}$ \\  
\hline
\hline
V$_{\rm red=1.00}$ &    9.8 &  2.6 / 0.1 &  0.1 & 0.2 & 0.1 / 0.01 & 0.01 & 0 & 0 & 0 & 0 & 0 & 0 \\
V$_{\rm red=0.80}$ &     22 &    8 / 0.3 &  0.4 & 1.2 & 0.1 / 0.01 & 0.01 & 0 & 0 & 0 & 0 & 0 & 0 \\
V$_{\rm red=0.75}$ &     72 &   29 / 1.1 &  1.2 & 1.7 & 0.1 / 0.01 & 0.01 & 0 & 0 & 0 & 0 & 0 & 0 \\
V$_{\rm red=0.60}$ &  4 550 & 1 066 / 41 & 40.7 & 2.6 & 0.1 / 0.01 & 0.01 & 0 & 0 & 0 & 0 & 0 & 0 \\
V$_{\rm red=0.50}$ & 14 534 & 1 984 / 76 & 73.5 & 3.9 & 0.2 / 0.7 & 0.03 & 0 & 0 & 0 & 0 & 0 & 0 \\
V$_{\rm red=0.40}$ & -- & -- & --               & 20  & 1.4 / 0.6 & 0.7 & 0 & 0 & 0 & 0 & 0 & 0 \\
V$_{\rm red=0.35}$ & -- & -- & --               & -- & -- & -- & 0 & 0 & 0 & 2 760 & 813 / 35 & 28.5 \\
V$_{\rm red=0.30}$ & -- & -- & --               & -- & -- & -- & 0.37 & 0.17 / 0.01 & 0.01 & 5 127 & 1 504 / 66 & 56.5 \\
V$_{\rm red=0.25}$ & -- & -- & --               & -- & -- & -- & 6.73  & 4.7 / 0.2 & 0.02 & 5 128 & 1 518 / 67 & 56.6 \\
V$_{\rm red=0.20}$ & -- & -- & --               & -- & -- & -- & 5 820 & 199 / 88 & 7.1  & -- & -- & -- \\
\hline
\end{tabular}
\tablefoot{
-- indicates that no calculations were carried out for these V$_{\rm red}$.}
\label{tab:ptcl}
\end{table*} 

\begin{figure}[htbp!]
\centering
\includegraphics[width=0.99\linewidth]{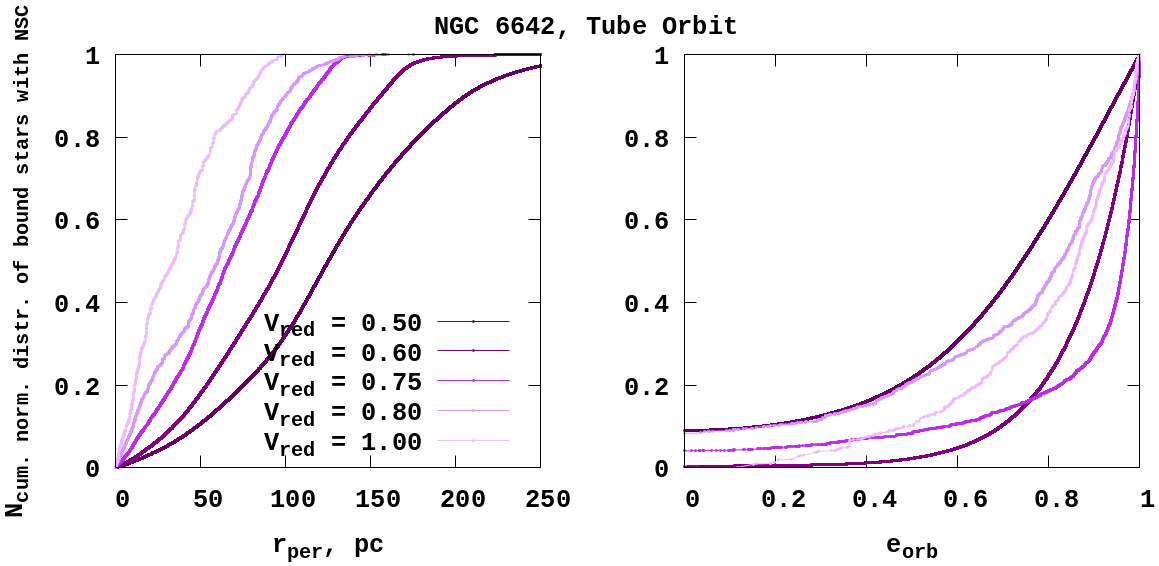}
\caption{Cumulative number distributions of the bound particles to the NSC as a function of the orbital stellar parameters $r_{\rm per}$ and $e_{\rm orb}$ with different velocity reduction factors for NGC~6642. Dark violet represents V$_{\rm red}$ = 0.5, and light violet represents V$_{\rm red}$ = 1.0.}
\label{fig:part-nsc}
\end{figure}

As an illustration for cluster NGC~6642, we present in Fig.~\ref{fig:part-nsc} the cumulative number distributions of the bound to NSC stars as a function of their orbital parameters: pericenter distance, $r_{\rm per}$, and orbital eccentricity $e_{\rm orb}$ with the selected $V_{\rm red}$ = 1.0 -- 0.5. The maximum pericenters of the stars are clearly shifted for the lower values of the reduction factors (shifting to higher kinetic energies). In the absence of the reduction factor ($V_{\rm red}$ = 1.0), the maximum $r_{\rm per}$ = 100 pc, as expected from the close-passage selection criteria for NSC. For a reduction factor equal to 0.5, the maximum pericenter shifted up to $\approx$ 250 pc.

\begin{figure}[htbp!]
\centering
\includegraphics[width=0.95\linewidth]{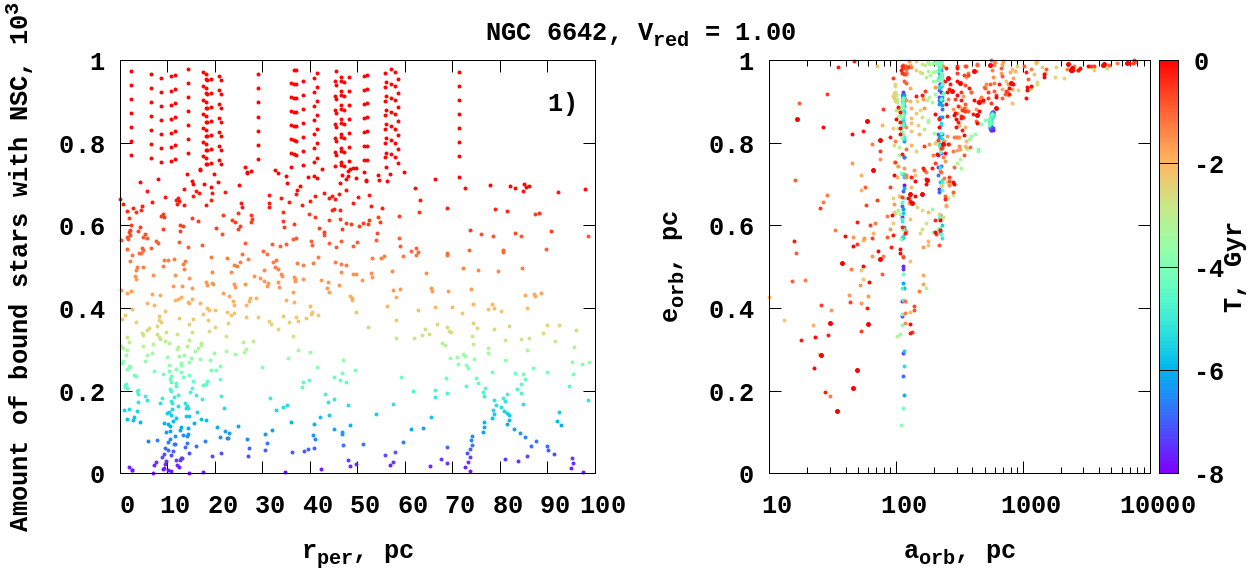}
\includegraphics[width=0.95\linewidth]{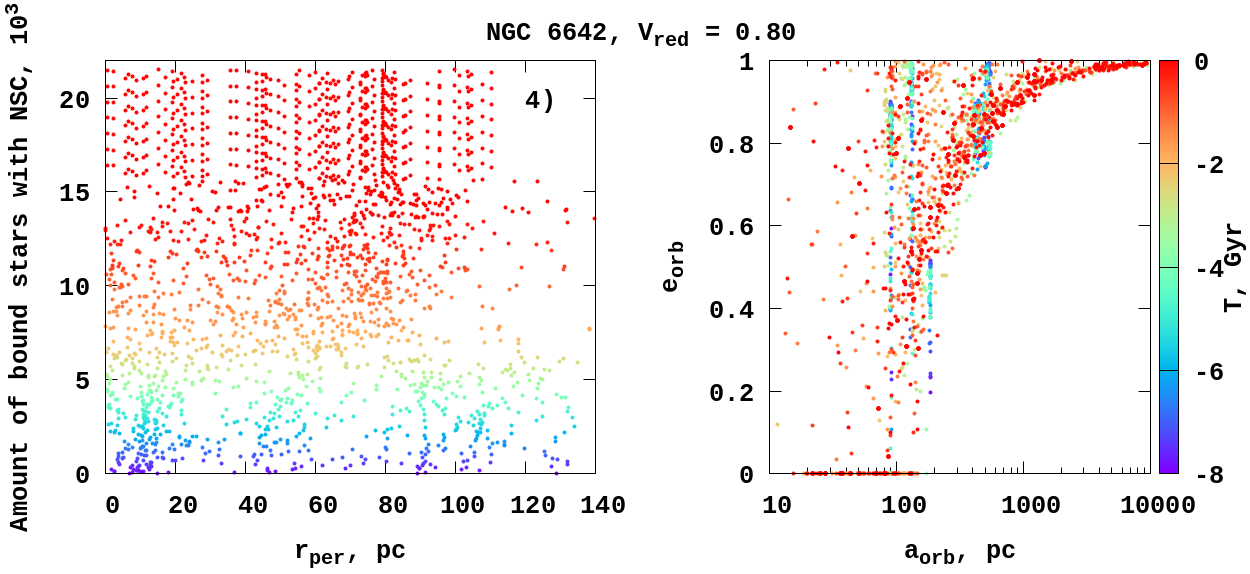}
\includegraphics[width=0.95\linewidth]{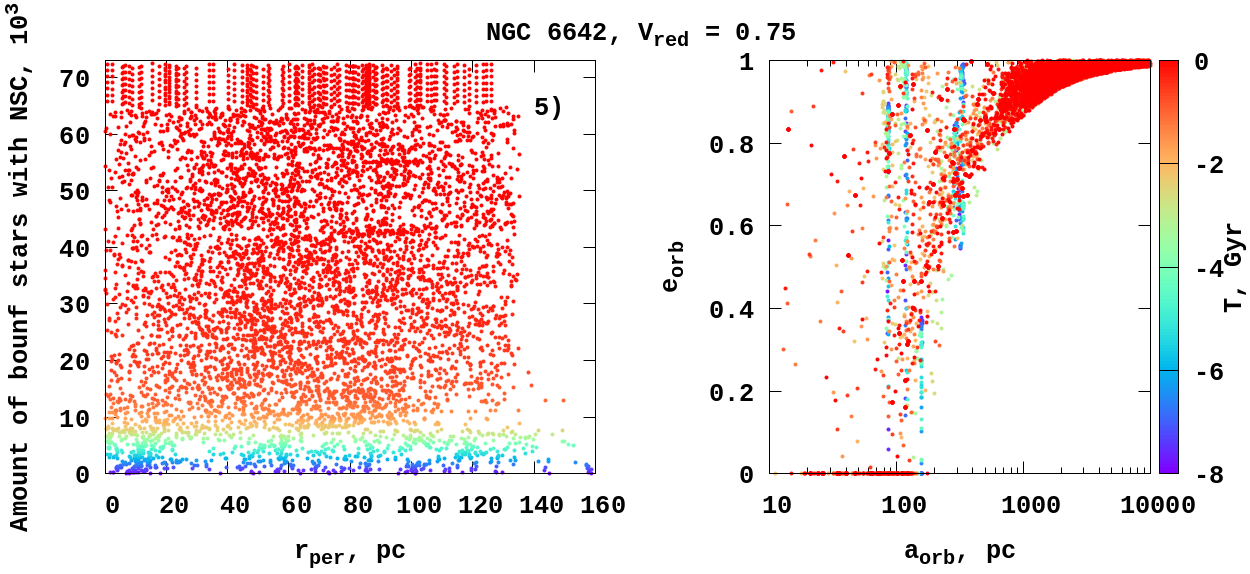}
\caption{Total bound number of stars as a function of their percienter for different reduction $V_{\rm red}$ = 1.0, 0.8, and 0.75. The orbital elements e$_{\rm orb}$ vs. a$_{\rm orb}$ for the stars are shown as a function of their GalC passage time.}
\label{fig:part-nsc-1}
\end{figure}

In Fig.~\ref{fig:part-nsc-1} (panels 1, 3, and 5) we show the total bound number of stars as a function of $r_{\rm per}$ colour-coded by the passage time. In the other panels (2, 4, and 6) we show the orbital element statistics ($e_{\rm orb}$ vs. $a_{\rm orb}$) for all bound particles for the different $V_{\rm red}$ = 1.0, 0.80, and 0.75. The cumulative distributions for different $V_{\rm red}$ show the clear effect of the maximum $a_{\rm orb}$ growth as a result of artificial velocity reduction in energy Eq. (see Eq.~\ref{eq:e_bound}). The plot shows that these bound particles have quite high orbital eccentricities ($\gtrsim0.8$), and as a result, quite large semimajor axes. We also can conclude that the most bound particles ($\approx80$\%) come close to the GalC in the last phase of integration time, starting from $\approx-2.0$~Gyr up to now (red).

NGC 6401, HP~1, and NGC~6681 present a completely different picture, see illustrations in Zenodo service\footnote{Total bound number of the stars with NSC for different velocity
redaction factor for NGC 6401, HP~1, and NGC~6681: \\~\url{https://zenodo.org/records/12107255}}. First of all, some bound particles only appear for quite low $V_{\rm red}$ values. An artefact of these low reduction velocity values is the quite distorted eccentricity distribution. The most bound particles have quite low eccentricities ($\lesssim$0.4). 

\begin{figure}[htbp!]
\centering
\includegraphics[width=0.75\linewidth]{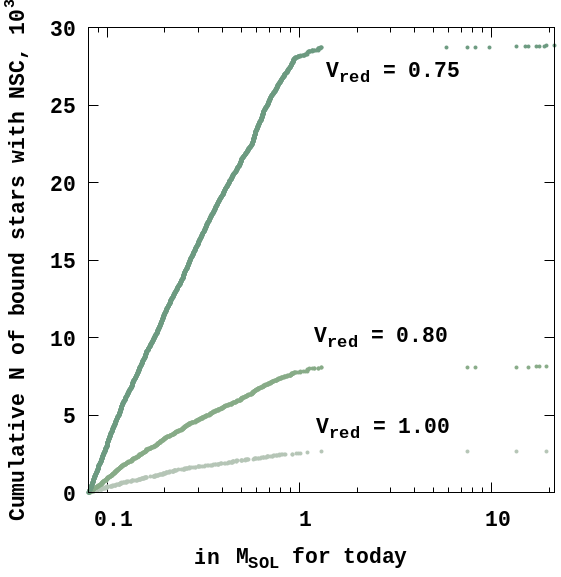}
\caption{Cumulative number of the NSC bound stars as a function of individual current stellar masses in M$_\odot$ for the NGC~6642 at the moment of passage near the GalC.}
\label{fig:cum-mass}
\end{figure}

Fig.~\ref{fig:cum-mass} presents an additional visualisation of the possible accumulated masses in the case of NGC~6642. The cumulative number distributions of the total bound NSC stars are shown as a function of the individual stellar masses today. Individual passages were only counted once. The plots show the strong influence of the IMF by the domination of the low-mass stars (different types of white dwarfs) in the total number of bound objects. For each $V_{\rm red}$, we calculated the total mass of NSC bound particles, which is also presented in Table~\ref{tab:ptcl}. We conclude that for the realistic $V_{\rm red}$ for NGC~6642, $V_{\rm red}$ = 1.0, 0.8, and 0.75 can potentially contribute to the NSC an amount of mass of about a few thousand M$_\odot$ and up to about ten thousand M$_\odot$.

\begin{figure}[ht]
\centering
\includegraphics[width=0.99\linewidth]{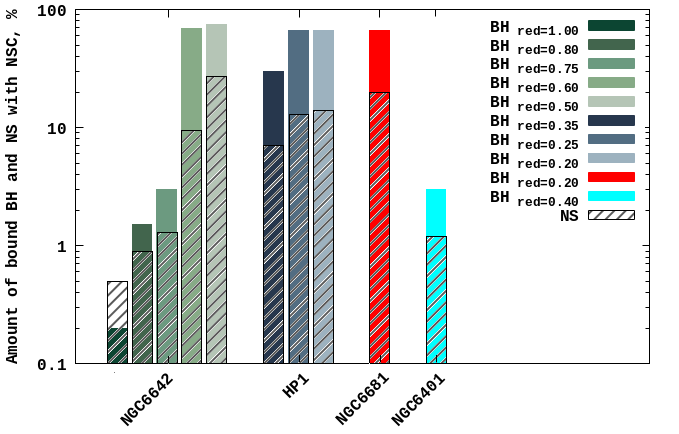}
\caption{Total relative number of compact stellar remnants that are bound with the NSC for NGC~6642, NGC~6401, HP~1, and NGC~6681.}
\label{fig:ns-bh-red-vel}
\end{figure}

In Fig.~\ref{fig:ns-bh-red-vel} we show the NSC bound fraction of high-mass remnants (BHs and NSs) for the total eight billion years of evolution. For the assumed realistic $V_{\rm red}$, we only obtain a few percent of bound remnants from the total number of high-mass remnants in the GC. Regardless of the always high NS formation kick velocities (in our stellar evolution prescription, the supernova fallback mechanism only works for the BH formation for some mass ranges; \citealt{Banerjee2020, Kamlah2022}), the total number of NSC bound NSs is comparable with the bound BH number. For a significant increase in the number of bound remnants, we need a much larger decrease in $V_{\rm red}$. This might for example be realised only in the presence of a significant gas fraction around the NSC. However, a proper description of the physical processes in such a complex and mixed dynamical environment is well beyond the scope of our current investigation.   

\section{Discussions and conclusions} \label{sec:disc-con}

We performed a set of direct $N$-body simulations of six GCs (NGC~6401, NGC~6642, Palomar~6, NGC~6681, NGC~6981, and HP~1) within a time-varying potential representing that of the Milky Way galaxy over an eight-billion-year period. These clusters were selected based on their potential interactions with the GalC. Incorporating stellar evolution, we analysed both the inner and global structure and evolution, covering aspects such as mass distribution, compact stellar remnants, and the overall evolution of the GCs. Furthermore, we quantified the stellar population loss from the GCs due to the GalC (hosting an SMBH), and the effect of this loss on to the NSC stellar content formation rate.

\subsection{Dynamical evolution of Galactic globular clusters} \label{subsec:concl-mass}

The orbital evolution of the globular star clusters in the Galactic tidal field on circular and not only circular orbits has already been studied in a few significant publications \cite{Baumgardt2003, Wang2016, Cai2016, GG2023, LM2024}. Our study is complementary to these works, but significantly differs in the sense of the time-variable external Galactic tidal field based on the IllustrisTNG-100 cosmological simulations (\cite{Pillepich2018}, \hyperlink{I23}{\color{blue}{Paper~I}} and \hyperlink{I23a}{\color{blue}{Paper~II}}). This time-variable nature of the external potential makes significant differences in the long-term dynamical evolution and mass loss of our selected globular star clusters. 

For the cluster survival in the strong tidal influence of the GalC, we generated our initial systems with a quite small half-mass radius (a few pc) and with a quite strong King concentration parameter (W$_0$ = 8 or 9; with r$_{\rm hm}$/r$_{\rm tid}$ $\approx$ 0.1; see Table \ref{tab:init-param}). This led to the highly concentrated GC models.

During the full eight billion years of evolution of the clusters, they lose from 70\% to 90\% of their initial masses. The internal structures of the clusters also change significantly. At the end of the simulation, the typical cluster concentration parameter $r_{\rm hm} / r_{\rm tid}$ reaches 11\% to 14\%. One of the exceptions is NGC~6981, which has a final  $r_{\rm hm} / r_{\rm tid}\approx4$\% due to the clusters long ($\approx15$~kpc) radial orbit in the halo region. 

The global parameters, such as the current tidal masses and half-mass radii of our selected clusters, age in relative agreement with their observed values. For example, based on the numerical simulations that we summarised in Table~\ref{tab:rhm-mass} (columns 2 and 5) versus the observed quantities in Table~\ref{tab:init-param} (columns 2 and 5), the relative current tidal mass mismatch for HP 1 is at a level of 1.4\% and at a half-mass radius of 4.6\%. For NGC~6401, the tidal mass differences are at 16\%, and for the radius, the difference is 30\%. The observational cluster parameters from the online catalogue\footnote{\url{https://people.smp.uq.edu.au/HolgerBaumgardt/globular/}} of \cite{Baumgardt2021} are constantly clarified, and as a result, the current tidal cluster radii and masses can significantly change over time.

As a result of our new stellar evolution prescription \citep{Banerjee2020, Kamlah2022MNRAS}, a significant fraction of NSs and BHs remnants lie in the GC. The total number of these high-mass remnants is $\approx0.84$\% from the GCs total initial particle numbers, according to our standard Kroupa IMF \citep{Kroupa2001}. Roughly 0.17\% of remnants are BHs. With our prescription, $\approx50$\% -- 60\% of this number are BHs with zero natal kicks, which forms a massive central BH subsystem. Inside this region lie the most massive BH remnants with masses from 30 to 40 M$_\odot$. The total mass of the BH subsystem can reach up to  $5 \times 10{^4}$ M$_\odot$. 

The global distribution of the GC tidal tails directly reflects the permanent mass loss of the clusters in the Galaxy tidal field. Generally, the cluster of unbound stars is distributed along the GC orbit. Due to the time-variable nature of our Galactic potential (coming from IllustrisTNG-100), the particle distribution slightly changes over the dynamical integration time during the last 8 Gyr. 

\subsection{Interaction with the Galactic nuclear star cluster} \label{subsec:concl-inter}

From the general point of view, GCs that can potentially lead to the process of NSC formation in our Galaxy have clearly already been destroyed by these long and dynamically violent processes \citep{Capuzzo-Dolcetta1993, Fabio2013, Minniti2021}. In this sense, our investigation of the GC survivors  can only show us a small fraction of the possible stellar accretion path to the formation of Galactic NSC. The GCs we studied are prototypes for the other (no longer visible) GCs that finally contribute the main mass to the future NSC of our Galaxy. 

All GCs we studied in detail, NGC 6401, NGC 6642, NGC 6681, and HP 1, exhibit significant mass loss of 70-90\% over a period of eight billion years. Stellar evolution accounts for about 30\% of the initial mass loss. Notably, NGC~6681 shows a lower mass-loss rate than its counterparts. Throughout the course of their evolution, these clusters develop a compact, nearly isothermal core. Specifically, NGC~6981 is distinguished by its extended unbound tidal tails.

The formation of stellar remnants significantly impacts the dynamics of GCs. BHs, representing approximately 0.17\% of the initial stellar population, tend to form centrally concentrated subsystems due to their high masses and weaker natal kicks. In contrast, NSs form in greater numbers (around 0.67\% of the initial stars), but exhibit higher kick velocities, leading to lower retention rates and a less centralised distribution within the clusters.

The fact that neutron stars are loosely bound to the clusters leads to their higher escape rates. This results in a more dispersed distribution of these objects throughout the Galaxy. However, because our selected clusters pass close to the GalC, many of these escaped neutron stars ultimately become concentrated in the central region. This spatial concentration around the GalC likely contributes to the excess of neutron stars in the central region of the Galaxy.

Based on our numerical modelling of the preselected real GC system dynamics in the time-variable Galactic potential on a cosmological timescale, we can conclude that only one of four of such systems might be a significant source of stars for the formation of Galactic NSC. Even in the worst-case scenario, that is, in the total absence of the dynamical friction in the GalC, about 2600 ($\approx0.1$\%) original stars ($\approx2.6$~million) from the cluster are accumulated into the MW NSC. In the assumption of the moderate dynamical friction (losing only 25\% of the orbital velocity in the process of possible future accretion) of the GC stars in the GalC stellar background, we already have 29 000 ($\approx1.1$\%) stars. 

NGC~6401 has a similar type of orbit and can provide an order-of-magnitude fewer potential stars for the NSC stellar population, even with the very optimistic scenario of  a reduced dynamical fraction velocity (a loss in orbital velocity of up to 40\%). 

Our other two candidates with different types of orbits (NGC~6681 with a radial orbit and HP~1 with an irregular orbit) contribute almost nothing to the potential NSC stellar population. Except for HP~1, which passes the GalC closely dozens of times in the last $\approx4$~Gyr and already has extended unbound tail structures, the potential contribution to the NSC is practically zero for any reasonable type of dynamical friction velocity loss (even up to 60\%). 

We analysed the mass distribution of the potential NSC stellar population in detail (see Fig.~\ref{fig:cum-mass}). We conclude that most of the potentially accreted stars are low-mass stars with masses below one M$_\odot$. The number of these stars is about 29000 for a realistic orbital velocity loss by 25\%.    

We analysed the stellar capture from the GC core or tail structure as as the main scenario of stellar accretion that forms NSC. Additionally, the process of collective interaction between the GalC and a few orbiting GCs can also be studied. This dynamical configuration allowed a wider range of potential captures due to the external perturbation (of the main cluster component and also of the tails) from the other fly-by GCs. A more detailed discussion of these collective interactions of GCs with the GalC is clearly beyond the scope of this paper. However, we have started a preliminary study of this dynamical investigation based on the orbital parameters of Terzan~2 and Terzan~4 \citep{Ishchenko2023c} and their interaction with the potential Galactic NSC (Ishchenko at al., in preparation). Our set of simulations already shows that a significant fraction of stars can be captured by the NCS if their limiting orbital velocities are below $\approx150-200$~km~s$^{-1}$. In the current study, the stars behave similarly ( Fig.~\ref{fig:red-vel}). Stars can only be captured into the forming NSC if the orbital velocity is below the limiting value (200~km~s$^{-1}$).    

All four GCs belong to the MW bulge component, but their orbital velocities are significantly different when they cross the GalC region (see Fig.~\ref{fig:red-vel}). Based on the figures, the significantly different fraction of stars can be understood as a source of the MW NSC for the different GCs from the sample we selected. From our simulations, we conclude that the GC stars captured by the NSC cannot be fully described based on the relative distance alone. The relative velocities between GC stars and NSC are clearly needed as an additional criterion.

In the case of NGC~6642, the number of high-mass remnants (BHs and NSs) that might can be captured by the NSC (for a realistic orbital velocity loss of 25\% through dynamical friction) reaches $\approx3$\% of the total number of high-mass remnants in the cluster. Originally, the cluster contained $\approx0.84$\% high-mass remnants. These analogues of the cluster mass-segregation process can cause the disproportionally high number of BH and NS remnants in the NSC stellar population.

\begin{acknowledgements}

The authors thank the anonymous referee for a very constructive report and suggestions that helped significantly improve the quality of the manuscript.

PB, MI, OV, MS and OS thanks the support from the special program of the Polish Academy of Sciences and the U.S. National Academy of Sciences under the Long-term program to support Ukrainian research teams grant No.~PAN.BFB.S.BWZ.329.022.2023.

This research has been funded by the Aerospace Committee of the Ministry of Digital Development, Innovations and Aerospace Industry of the Republic of Kazakhstan, Grant No. BR20381077.

MS acknowledges the support under the Fellowship of the President of Ukraine for young scientists 2022-2024.

\end{acknowledgements}

\bibliographystyle{mnras}  
\bibliography{gc-mass-evol}   

\begin{appendix}

\section{Overview of the selected GCs}\label{app:gc-lit}

\subsection{NGC~6681}\label{subsec:ngc_6681}
NGC~6681 or M~70 was first discovered by W. Herschel in the 18th century and is located in the Sagittarius constellation. Based on preliminary research, its age is estimated to be around 11.6~Gyr, as indicated by \cite{Forbes2010} and, according to the \cite{Malhan2022}, is believed to have originated within the Galaxy. 

Our numerical orbital modelling results reveal that over the 8~Gyr the $r_{\rm apo}$ has 6.6~kpc and for $r_{\rm peri}$ -- 270 pc in time evolving potential. NGC~6681 has a long radial orbital shape with an anti-clockwise rotation. During one orbital turn, the GC passes through various regions of the Galaxy, including the halo (inner part), and the thick and thin disk, and has close passages at the distance of $\approx42$~pc near the GalC in the halo. As illustrated in Fig.~\ref{fig:gc-coll}, NGC~6681 also exhibits a low probability of direct collisions in $N_{4r_{\rm hm}}$ and $N_{2r_{\rm hm}}$, relative to its own half-mass radii, throughout its orbital evolution.

According to the \cite{Franch2009}, NGC~6681 is associated with a population of old clusters, and its luminosity values [Fe/H] are valuated from $-1.3$ to $-1.63\pm0.07$~dex \citep[see papers][]{Harris1996,Boyles2011,OMalley2017}.
                        
\subsection{NGC~6642}\label{subsec:ngc_6642}
NGC~6642 is a GC with the type IV stars in the Sagittarius constellation, which was discovered by W.~Herschel in the 18th century. This cluster is considered one of the oldest known GCs. According to \cite{Balbinot2009} its preliminary age is estimated to be around 13.8~Gyrs, and in line with \cite{Malhan2022} it is believed to have originated within the Galaxy. 

According to our orbital numerical modelling results, during the 8~Gyr the $r_{\rm apo}$ has 1.7 kpc and in  $r_{\rm peri}$ -- 400~pc in time evolving potential. NGC~6642 have a tube-shaped orbit with clockwise rotation. Studies conducted by \cite{Boyles2011} and \cite{Balbinot2009} for NGC~6642 suggest luminosity values [Fe/H] are valuated from $-1.25$ to $-1.80\pm0.2$~dex.

The observed present-day mass function in NGC~6642 indicates significant mass loss of low-mass stars and that the cluster has spent more than $\approx90$\% of its lifetime, as evidenced by strong tidal effects \citep{Balbinot2009}. NGC~6642 has a very dense central core and is probably composed of stars that have undergone (or are undergoing) mergers. The inability to clearly identify the radial density profile of the core supports the idea that NGC~6642 have a collapsed core.

\cite{Balbinot2009} analysed the luminosity function and present-day mass function with their variation as a function of radius and found the evidence for mass segregation, especially in the range $0.4 \leqslant m/{\rm M_{\odot}} \leqslant 0.8$. In NGC~6642 a decrease in the number of stars leads to a decrease in luminosity and mass, a phenomenon more pronounced in its central regions. This inversion in the present-day mass function slope, while atypical, has also been confirmed in other GCs through dynamical $N$-body simulations, as seen in works by \cite{Andreuzzi2001}, \cite{Marchi2007}, and \cite{Paust2009}. The authors explained it in terms of disk and bulge which shocked NGC~6642 when it has passages near the GalC, rendering it highly susceptible to tidal disruption \citep[see also][]{Baumgardt2003}. 

\subsection{GC NGC~6981}\label{subsec:ngc_6981}
NGC~6981 or M~72 is a GC in the Aquarius constellation discovered by Pierre Méchain in the 18th century. This cluster is associated with Gaia-Sausage (Enceladus) merger event \citep{Malhan2022}.

NGC~6981 is a weakly metallic cluster with [Fe/H]$=-1.48\pm0.03$~dex (RR Lyrae stars Fourier light-curve decomposition method, \citealt{Bramich2011}). The age is ambiguously estimated: $\approx9.5$~Gyr by \cite{Meissner2006} or $\approx12.75$~Gyr by \cite{VandenBergh2011}.

The angular diameter of the cluster is 5.9~arcmin \citep{Dreyer1988}, while the linear diameter is $\approx32$~pc \citep{Alcaino1977}. The orbit of NGC~6981 is retrograde \citep{Zinn1985}. The cluster own rotation is low, so it has $b/a$ flattening ratio close to unity. Our previous numerical orbital modelling showed that $r_{\rm apo}$ has 24.7 kpc and $r_{\rm peri}$ --76 pc over 8~Gyr. Other authors have various views on the $r_{\rm peri}$ value. For example, \cite{Balbinot2018} estimated $r_{\rm peri}$ of the GC orbit to be $10.66\pm0.46$~kpc with the eccentricity $e = 0.99\pm0.01$, which indicates a very elongated orbit. However, \cite{Baumgardt2019} gives severely different values: $r_{\rm peri} = 1.29\pm0.74$~kpc, $r_{\rm apo} = 24.01\pm4.52$~kpc. 

NGC~6981 total mass is estimated also differently: $1.68\times10^{5}\rm\;M_{\odot}$ \citep{Boyles2011} or $8.1\times10^{4}\rm\;M_{\odot}$ \citep{Baumgardt2019}. \cite{Balbinot2018} obtained the initial mass of the GC: $M_{\rm ini} = (2.12\pm0.09)\times10^{5}\rm\;M_{\odot}$. However, \cite{Piatti2019} believe that the value of the mass lost by the NGC~6981 during tidal disruption is $\approx40\%$ of the $M_{\rm ini}$ of the cluster. Taking into account the mass loss of the cluster through evolutionary effects, it turns out that the observed mass of the cluster is only $9\%$ from $M_{\rm ini}$ \citep{Piatti2019}.

There are many variable stars in NGC~6981. \cite{Ji2013} showed that the proportion of binary stars decreases with increasing distance to the centre. NGC~6981 has extra-tidal structures, which confirms its ``accreted cluster'' nature. \cite{Grillmair1995}, using photographic photometry, did not find emphasised tidal tails around the cluster. However, the study of \cite{Piatti2021} showed the presence of structures in the form of a non-rounded halo and debris along the tail. In order to understand the spatial distribution of extra-tidal debris in NGC~6981, the authors simulated the dynamics of the cluster with an initial King profile \citep{King1966} and a mass of $2\times10^{5}\rm\;M_{\odot}$ up to two billion years in the Galaxy three-component potential (bulge, disk, halo). Modelling was done using the GADGET-2 code \citep{Springel2005}. As a result of the simulation, an extended trail of destroyed stars was observed, lying parallel to the direction of the cluster velocity vector. Thus, NGC~6981 significantly lost mass during its passage through the pericenter of the orbit. Also \cite{Zhang2022} suggest a low-luminosity extended envelope-type structure around the cluster.

The study by \cite{askar2017mocca-mukha, Askar2018} utilised the MOCCA  simulations to predict the presence of BHs, binary BHs, and BH sub-systems in the cluster. They also give the corresponding approximate numbers for the  NGC~6981 cluster. They suggest that NGC-6981 could host black holes (total number of BHs in GC: $N=135^{+75}_{-44}$), including binary black holes: $N=8^{+17}_{-5}$ (involving other stars), and $84^{+37}_{-22}$ BHs in a black hole sub-system. These numbers we should compare with our numbers with big care, due to the limitations of the MOCCA survey and the approximations used to derive the BH numbers in individual clusters \cite[][see Eq.~(1) and Fig.~1]{Askar2018}. 

\subsection{Palomar~6}\label{subsec:pal_6}
GC Palomar~6 was first discovered by R.~G.~Harrington and F.~Zwicky in the 1955 and is located in the Ophiuchus constellation \citep{Abell1955}. The authors in \cite{Souza2021} indicate that the Palomar~6 is located in the bulge and that it was probably formed in the bulge in the early stages of the MW formation, sharing the chemical properties with the family of intermediate metallicity very old clusters such as M~62, NGC~6522, NGC~6558, and HP~1. The best distance, reddening, and well-constrained metallicity values provided us with the first derivation of age for Palomar~6 as $12.4\pm0.9$~Gyr, therefore it is among the oldest GCs in the Galaxy.

According to our orbital numerical modelling results in time variable potential, during the 8~Gyr the $r_{\rm apo}$ has 3.4 kpc and for $r_{\rm peri}$ -- 150 pc. Palomar~6 has tube-shaped orbit with clockwise rotation. 

The controversy on which Galactic component Palomar~6 is part of is also due to an uncertain metallicity. The metallicity estimations of this object valuated from [Fe/H] $\approx$-1.30 to -1.08~dex \citep{Ortolani1995, Lee2004}. Estimations were done based on a reddening-free index based on data observed at the ESO NTT-EMMI by the slope of the red giant branch and the presence of a red horizontal branch. Also, by analysing the slope of the RGB on the near-infrared CMD with NICMOS3 JHK bands. Spectroscopic analysis from high-resolution NIR spectra of three RGB stars by the same authors resulted in [Fe/H]$=-1.08\pm0.06$~dex. According to the newest observations, a metallicity is [Fe/H]$=-1.1\pm0.09$~dex and was confirmed from a high-resolution spectroscopic analysis of five probable member stars observed with the CSHELL spectrograph at the NASA Infrared Telescope Facility \citep{Souza2021, Boyles2011}.

\subsection{GC NGC~6401}\label{subsec:ngc_64010}

NGC~6401 (also known as GCL1735-238 and ESO520-SC11) is a GC in the constellation Ophiuchus, discovered by William Herschel in 1784. The GC in the list of clusters within a box of $20\degr\times20\degr$ around the GalC \citep{barbuy1999metal-mukha}.
The NGC~6401 cluster has low metallicity, with [Fe/H] ranging from -1.12 \citep{davidge2001near-mukha} to -1.37 dex \citep{valenti2007near-mukha}. However, \cite{santos2004ages-mukha} reports that the cluster has a metallicity of [Fe/H]=-0.96 and an age of 11.8~Gyr. 

Also, 2010 updated Harris catalogue \citep{Harris2010} shows that NGC~6401 has [Fe/H] = -1.02~dex, an absolute magnitude of -7.90, and a heliocentric radial velocity of $-65.0\pm8.6$~km~s$^{-1}$ (radial velocity relative to the Solar neighbourhood LSR is -55.3~km~s$^{-1}$). Additionally, NGC~6401 contains variable stars. \cite{tsapras2016variable-mukha} found 34 RR~Lyrae (23~RRab and 11~RRc) stars, and their work also determined that NGC~6401 is an Oosterhoff type I cluster (the cluster has formed early in the history of the MW galaxy).

The NGC~6401 is a buldge GC with an angular size 0.35~arcmin and in this area was found 93~stars \citep{chen2010morphological-mukha}. The cluster's apocenter at $r_{\rm apo}$ = 3.5~kpc and pericenter at $r_{\rm peri}$ = 220 pc according to the our numerical simulations.

The study by \cite{askar2017mocca-mukha, Askar2018} utilised the MOCCA  simulations to predict the presence of BHs, binary BHs, and BH sub-systems in the cluster. They also give the corresponding approximate numbers for the  NGC~6401 cluster. It was found that there are $121^{+65}_{-38}$ BHs, $8^{+16}_{-5}$ of them are BH binary systems (involving other stars as a second component), and potentially a BH sub-system containing $73^{+30}_{-19}$ BHs. These numbers we should compare with our numbers with big care, due to the limitations of the MOCCA survey and the approximations used to derive the BH numbers in individual clusters \cite[][see Eq.~(1) and Fig.~1]{Askar2018}. 

\subsection{HP 1}\label{subsec:hp_1}
GC HP~1 was discovered by \cite{Dufay1954} in the Haute-Prevonce Observatory. HP~1 is located in Ophiuchus and has an apparent magnitude of 11.59 and a diameter of 1.2~arcmin. According to our numerical simulations shows that $r_{\rm apo}$ -- 2.4~kpc and $r_{\rm peri}$ has 850 pc for over 8 Gyr.

HP~1 is an $\alpha$-enrichted metal-poor GC with a blue horizontal branch \citep{Kerber2019}. \cite{Ortolani1997} deduced a metallicity of [Fe/H] $\approx-1.0$~dex from color-magnitude diagram. \cite{Dias2016} obtained value of [Fe/H] = -1.17$\pm{0.07}$~dex from low-resolution spectroscopy of eight red giants. \cite{Barbuy2016} carried out a detailed analysis of 8~stars and got values of [Fe/H] = -1.06$\pm{0.15}$~dex. Low [Fe/H] values and a blue horizontal branch show that HP~1 is a very old GC. Statistical isochrone fits in the Near Infrared photometry and optical-NIR colour-magnitude diagrams indicating an age of 12.8~Gyr \citep{Kerber2019}. This fact confirms that HP~1 is one of the oldest clusters in the MW. 
\cite{Ortolani1997}, \cite{VandenBergh2011} proposed that HP~1 could be a halo cluster that is presently passing through the Galactic bulge, or 
be an object that has been partly disrupted by tidal forces. \cite{Ortolani2011} numerical simulations show that HP~1 does not wade into the halo and is confined within the Galactic bulge.  
\cite{Barbuy2016} found that there are some stars with slightly lower metallicity and lower radial velocities. Differences in metallicity might show that there are multiple stellar generations in a cluster. Authors propose that HP~1 is probably influenced by tidal effects that lead to possible perturbations or even disruption.

\clearpage


\section{NS global distribution in the Galaxy}\label{app:NS-distr}

We have the galactocentric coordinates ($x_{\ast},\ y_{\ast},\ z_{\ast}$) of the NS from simulations. To transform these coordinates into Galactic coordinates (galactic latitude and longitude), we follow a several-step procedure from \cite{kalambay2022mock-mukha}. Initially, we utilise the galactocentric coordinates of the Sun: $X_{\odot}= -8178 \, \text{pc}$ \citep{Gravity2019}, $Y_{\odot} = 0 \, \text{pc},\ Z_{\odot} = 20\, \text{pc}$ \citep{Bennett2019}. This enables us to determine the heliocentric coordinates of the NS, as described in Eq.~(\ref{eq:heliocentric}). Then, we find the heliocentric distance and subsequently, we convert to Galactic coordinates using Eq.~(\ref{eq:heliodis})
\begin{equation}
\begin{gathered}
X_{\ast} = x_{\ast} - X_{\odot},\ \ \ \ \ \ \
Y_{\ast} = y_{\ast} - Y_{\odot}, \ \ \ \ \ \ \ 
Z_{\ast} = z_{\ast} - Z_{\odot} 
\end{gathered}
\label{eq:heliocentric}
\end{equation}
\vspace{-2.0\baselineskip} 
\begin{equation}
\begin{gathered}
d = \sqrt{X_{\ast}^2 + Y_{\ast}^2 + Z_{\ast}^2},\
l = \arctan\left(\frac{Y_{\ast}}{X_{\ast}}\right),\  
b = \arcsin\left(\frac{Z_{\ast}}{d}\right)
\end{gathered}
\label{eq:heliodis}
\end{equation}

\begin{figure}[htbp!]
\centering
\includegraphics[width=0.95\linewidth]{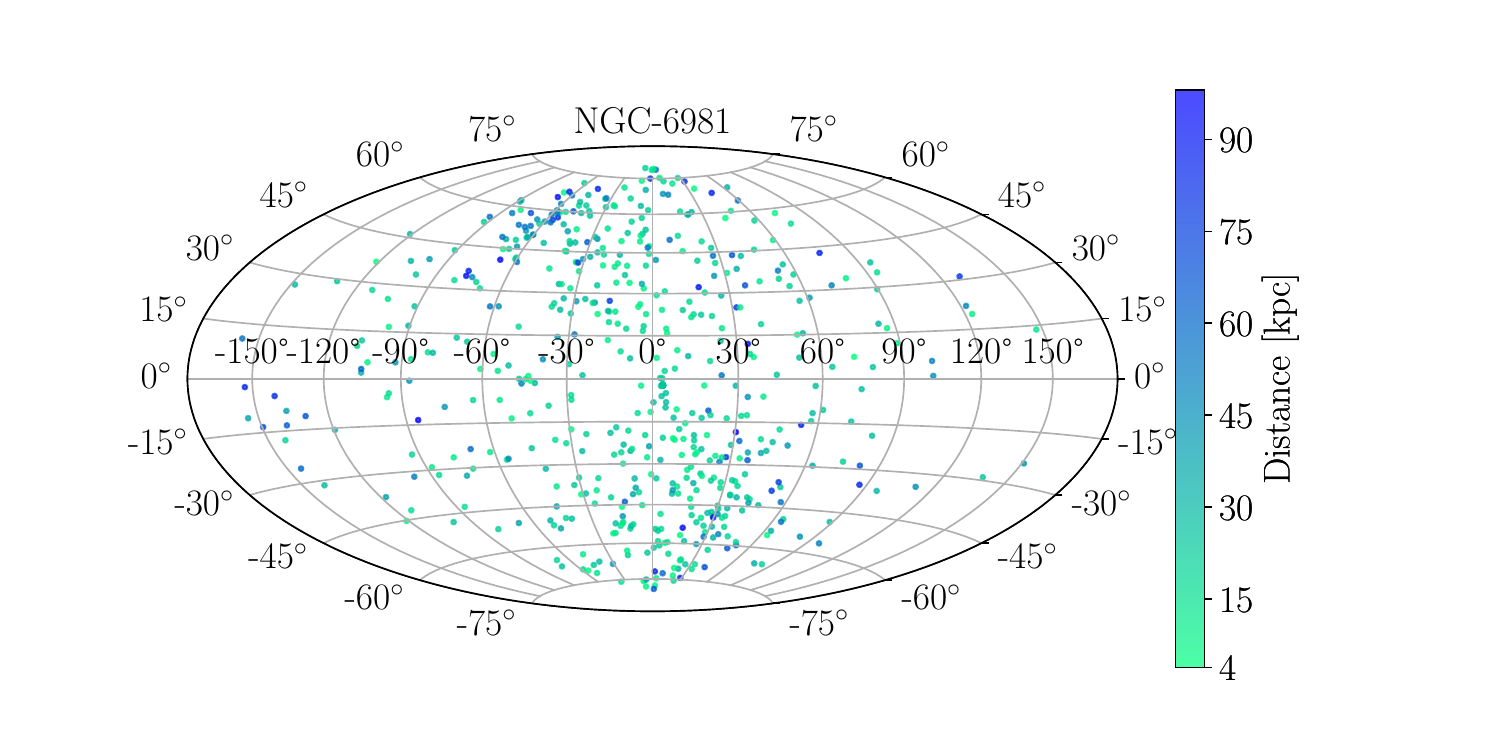}
\caption{Mapping the NSs Galactic coordinates ($l$ and $b$) in Hammer-Aitoff projection for NGC~6981, based on numerical simulation at present day. The colour bar shows the transitions from light green at close distances to blue at far distances, up to 100~kpc from the Sun.}
\label{fig:hammer-6981}
\end{figure}

\begin{figure}[htbp!]
\centering
\includegraphics[width=0.95\linewidth]{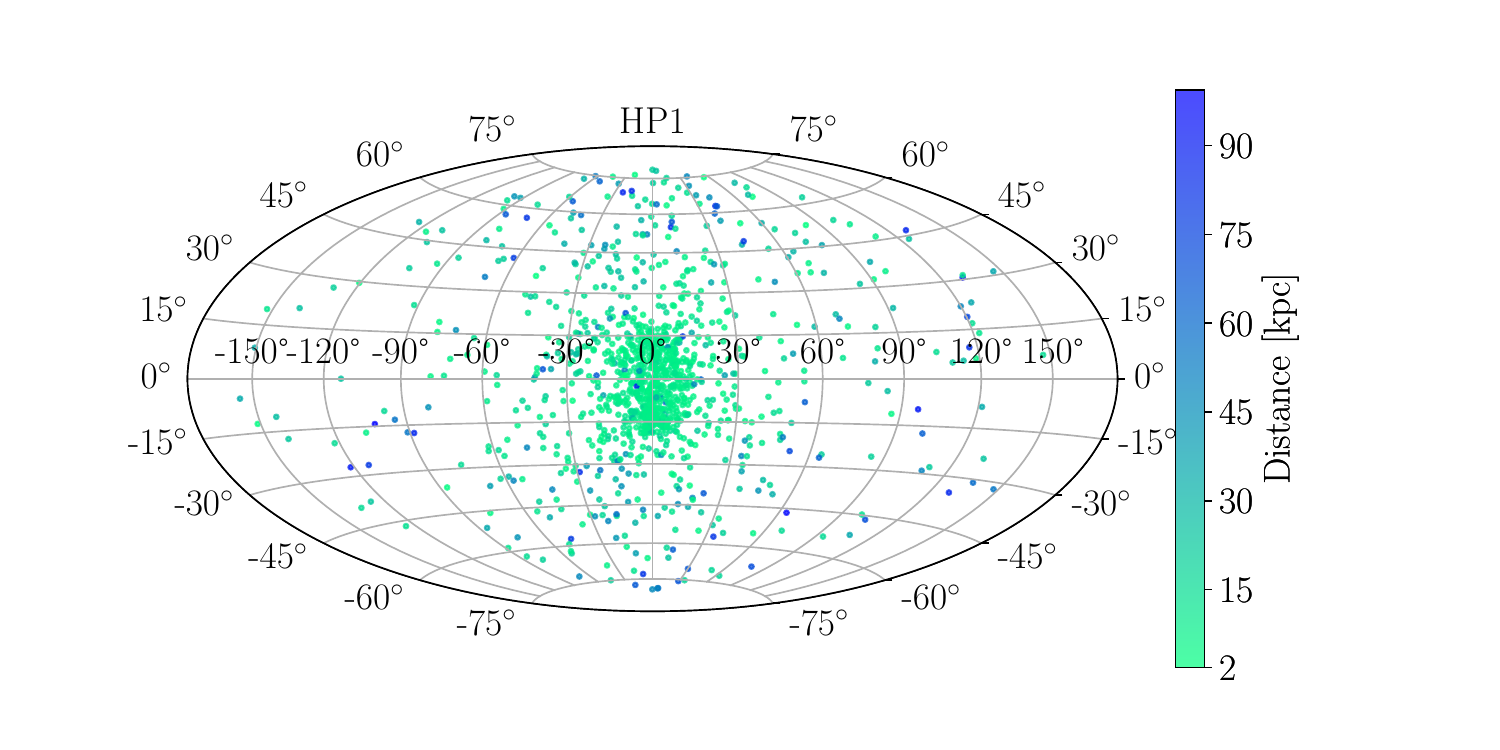}
\caption{Same as  Fig.~\ref{fig:hammer-6981} but for GC HP~1. }
\label{fig:hammer-hp1}
\end{figure}

\begin{figure}[htbp!]
\centering
\includegraphics[width=0.95\linewidth]{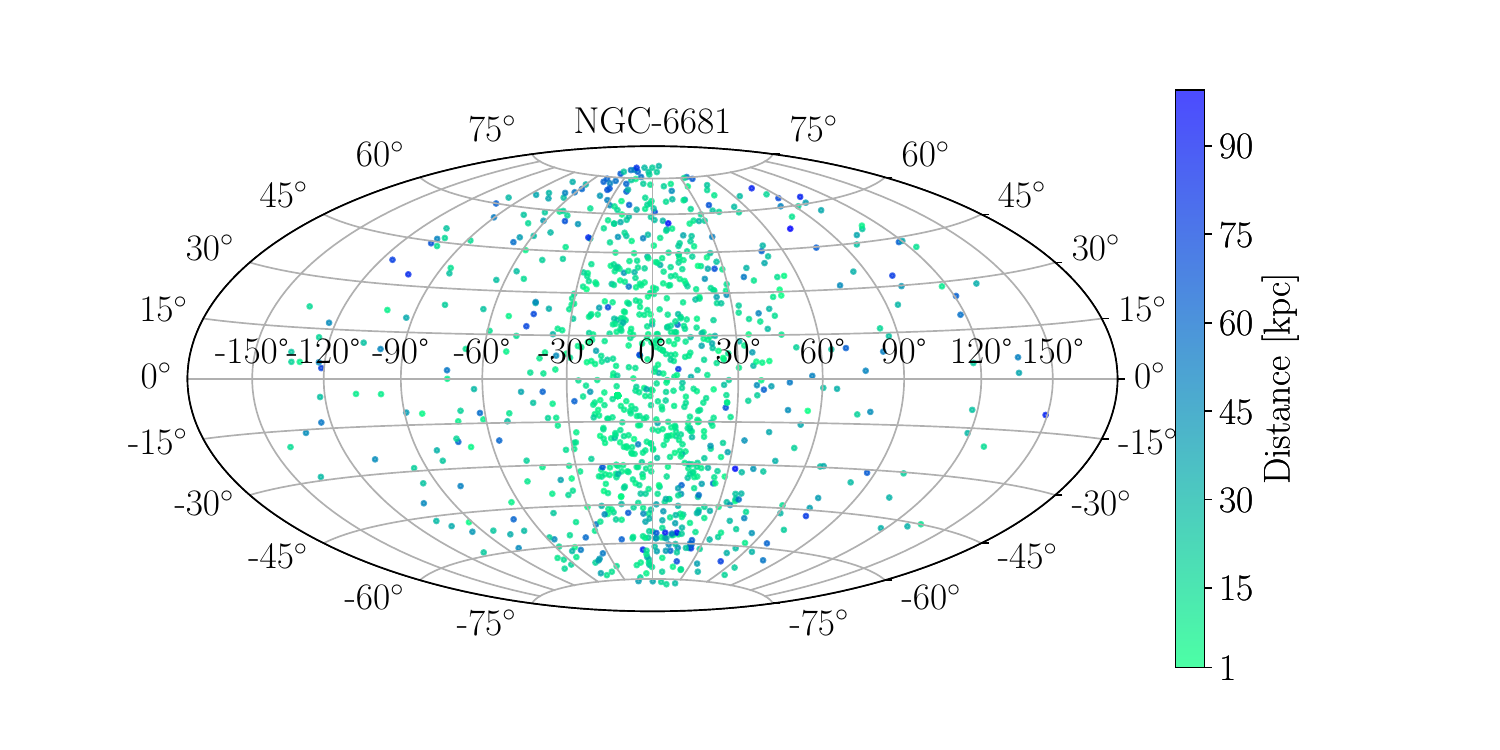}
\caption{Same as  Fig.~\ref{fig:hammer-6981} but  for GC NGC~6681. }
\label{fig:hammer-6681}
\end{figure}

\begin{figure}[htbp!]
\centering
\includegraphics[width=0.95\linewidth]{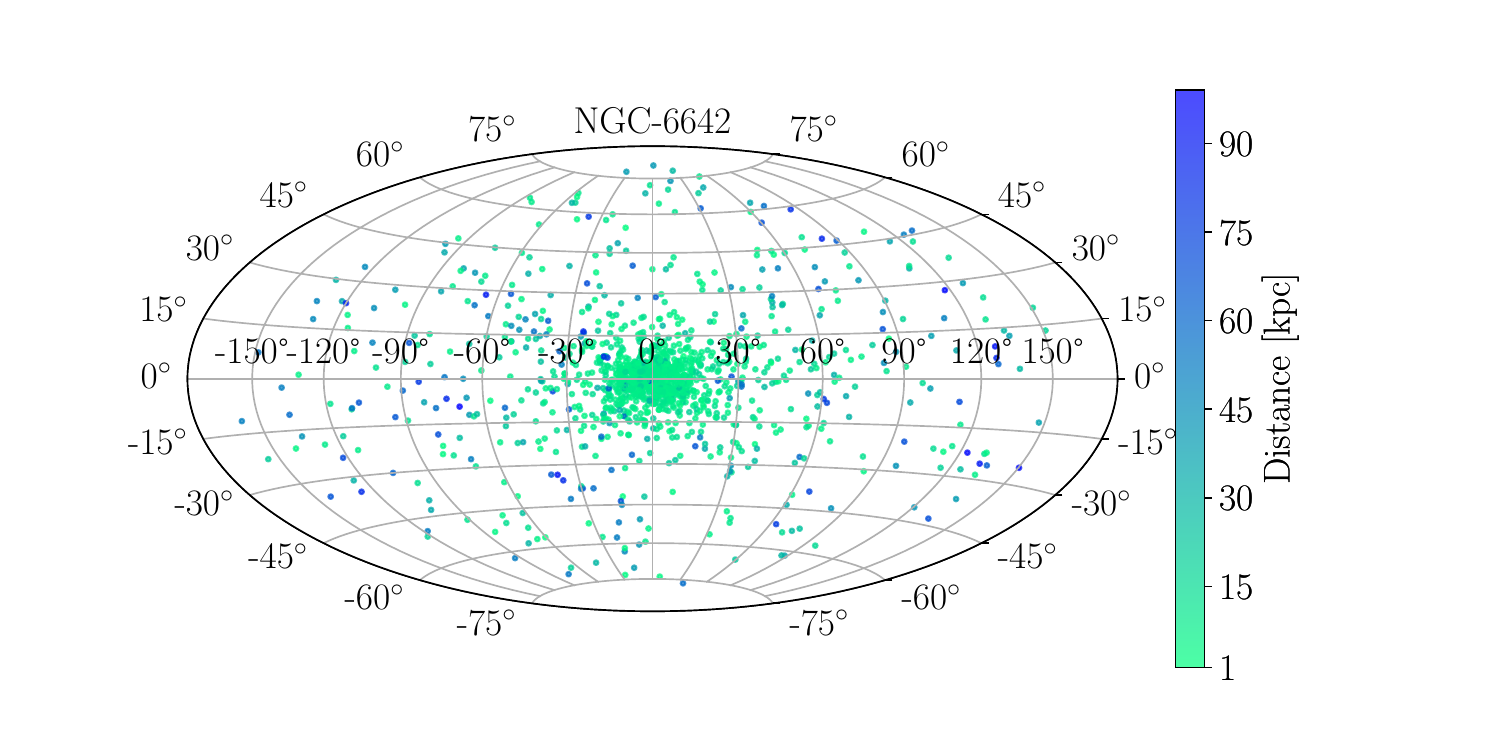}
\caption{Same as  Fig.~\ref{fig:hammer-6981} but  for GC NGC~6642. }
\label{fig:hammer-6642}
\end{figure}

\begin{figure}[htbp!]
\centering
\includegraphics[width=0.95\linewidth]{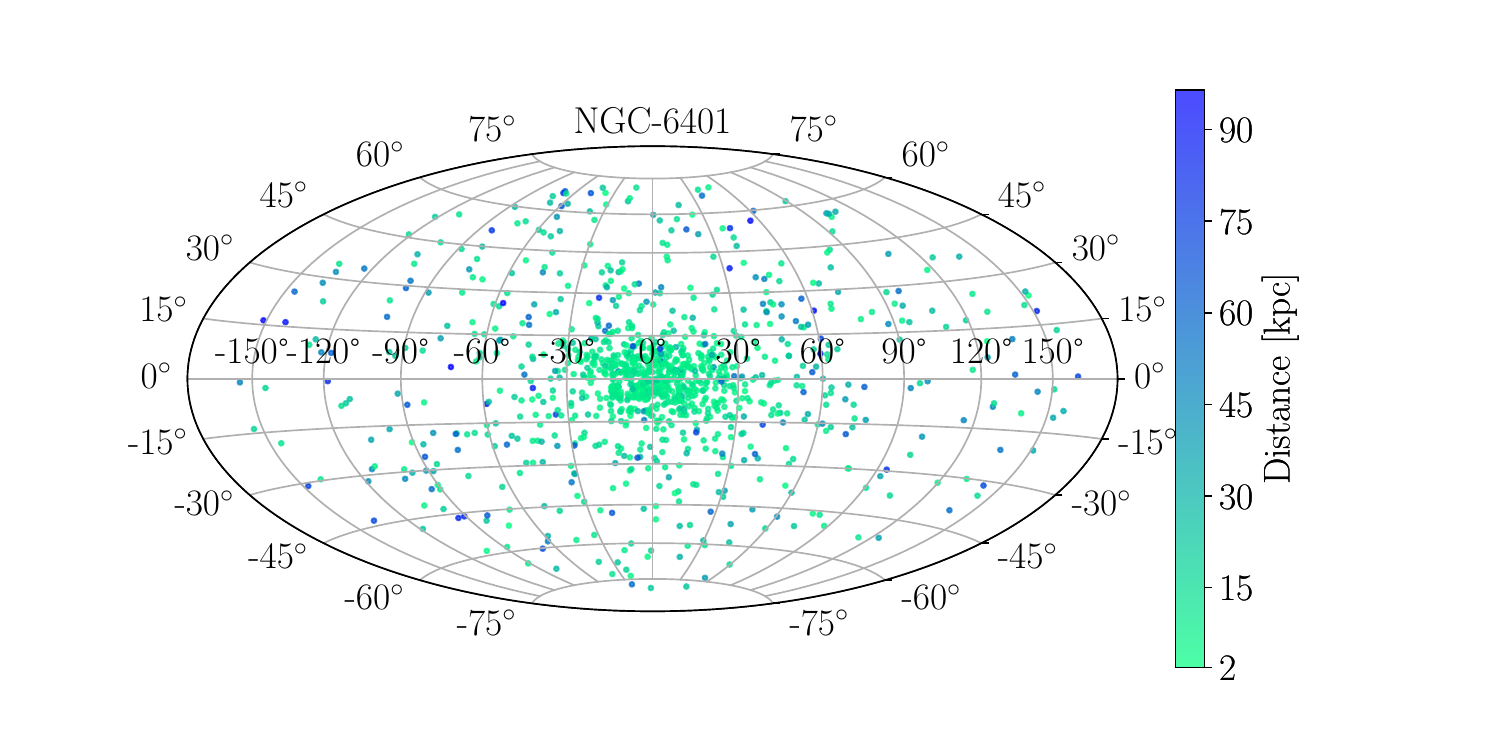}
\caption{Same as  Fig.~\ref{fig:hammer-6981} but for GC NGC~6401. }
\label{fig:hammer-6401}
\end{figure}

\begin{figure}[htbp!]
\centering
\includegraphics[width=0.95\linewidth]{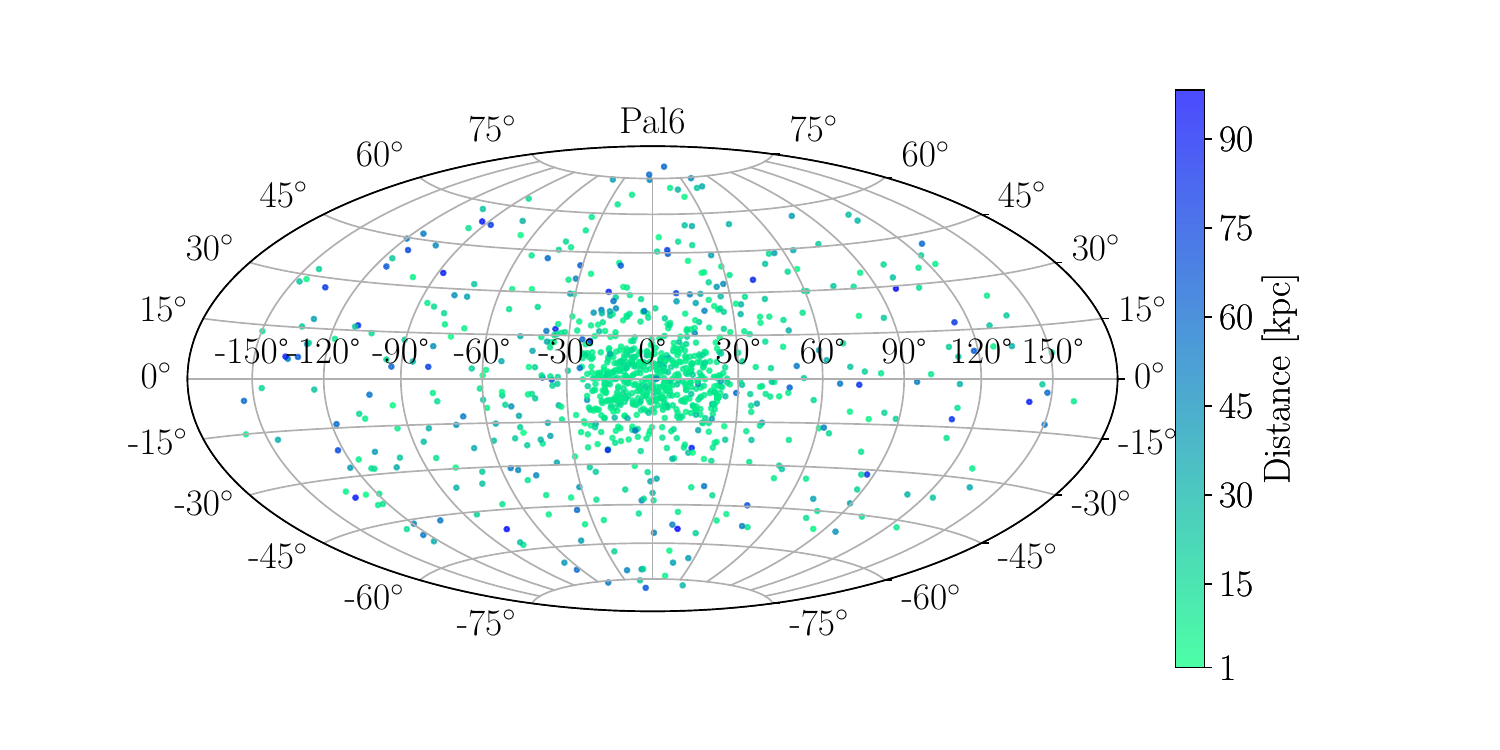}
\caption{Same as  Fig.~\ref{fig:hammer-6981} but  for GC Palomar~6. }
\label{fig:hammer-pal6}
\end{figure}

We used the standard Hammer-Aitoff (H-A) equal-area projection\footnote{\\~\url{ https://en.wikipedia.org/wiki/Hammer_projection}} to visualise Galactic coordinates, as illustrated in Figs.~\ref{fig:hammer-6981}--\ref{fig:hammer-pal6}. The conversion from galactic coordinates ($l$, $b$) to the H-A projection ($x_{\rm H-A}$, $y_{\rm H-A}$) was accomplished using the following equations:
\begin{equation}
\begin{aligned}
x_{\rm H-A} &= \frac{2\sqrt{2} \cos{b} \sin{(l/2)}}{\alpha},\ \ \ \ \ \  y_{\rm H-A} &= \frac{\sqrt{2}\sin{b}}{\alpha}, \\
\end{aligned}
\label{eq:hammer-aitoff}
\end{equation}
where $\alpha$ is defined as $\alpha=\sqrt{1+ \cos{b} \cos{(l/2)}}$. This projection allows for a comprehensive representation of celestial coordinates in a two-dimensional plane.


\clearpage

\section{Selected GCs global stellar density distributions in the Galactic coordinates }\label{app:star-dist}
\begin{figure*}[ht]
\centering
\includegraphics[width=0.99\linewidth]{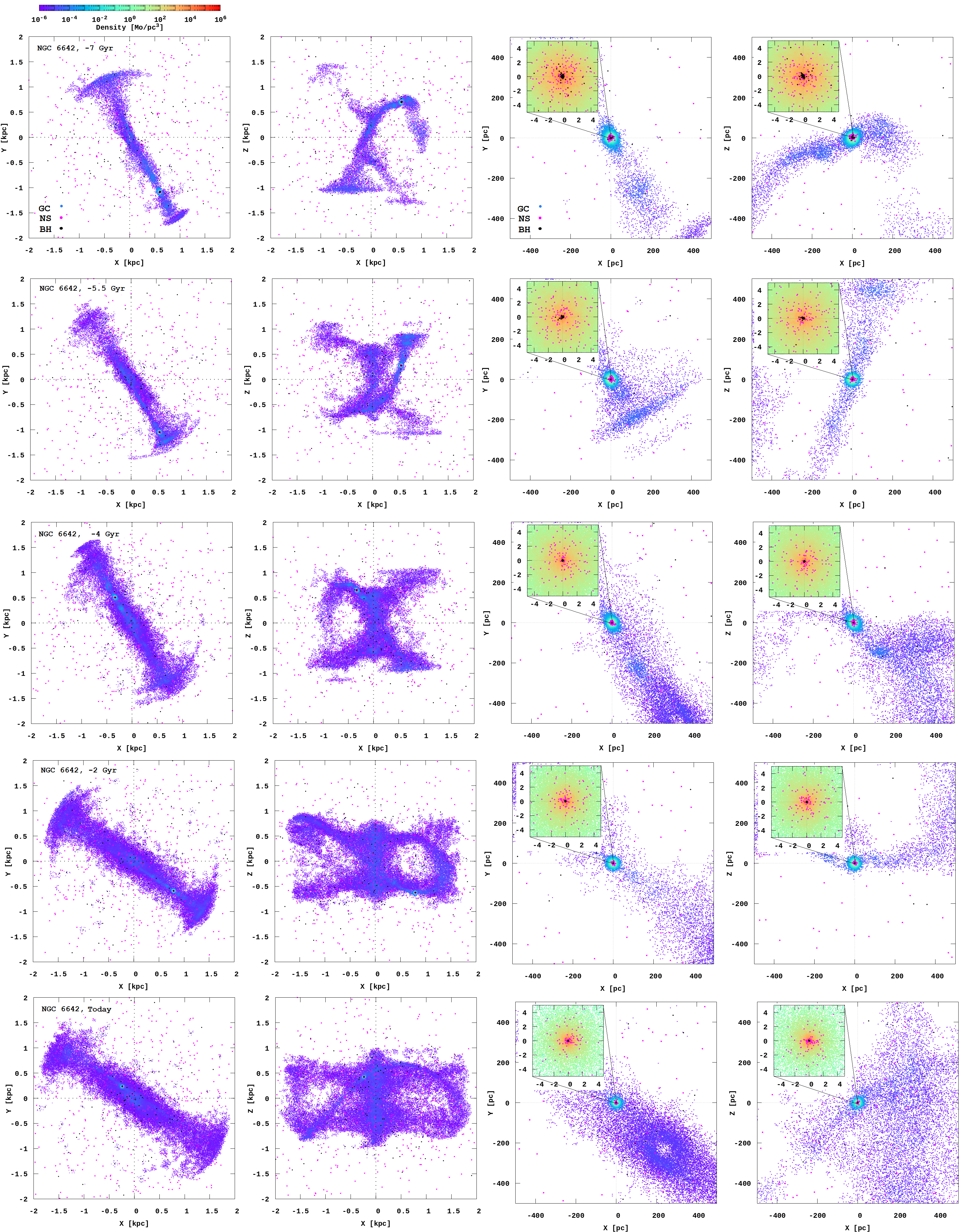}
\caption{NGC~6642 clusters density distributions in {\tt 411321} TNG-TVP external potential. The orbital global evolution is present in two projections ($X$, $Y$) and ($X$, $Z$), \textit{two left panels}. GC central part in local frame with the BHs (black dots), NSs (magenta dots) and with detail central area (with box size 10 pc) we present in \textit{two right panels}. The total time of integration is 8~Gyr forward time.}
\label{fig:ngc_6642_star_loss}
\end{figure*}

\clearpage

\begin{figure*}[ht]
\centering
\includegraphics[width=0.99\linewidth]{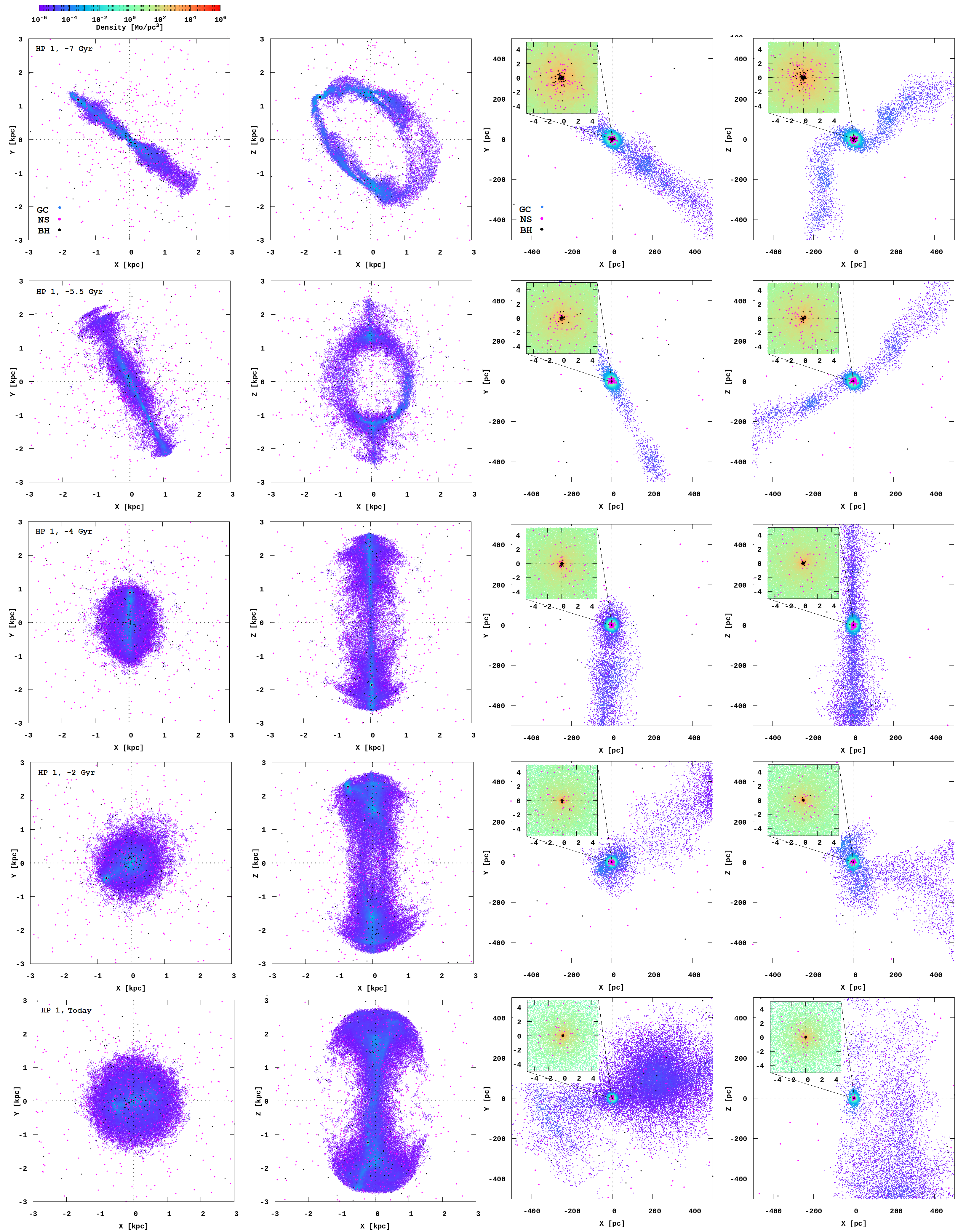}
\caption{Same as in Fig.~\ref{fig:ngc_6642_star_loss} for GC HP~1.}
\label{fig:hp1_star_loss}
\end{figure*}

\clearpage

\begin{figure*}[ht]
\centering
\includegraphics[width=0.99\linewidth]{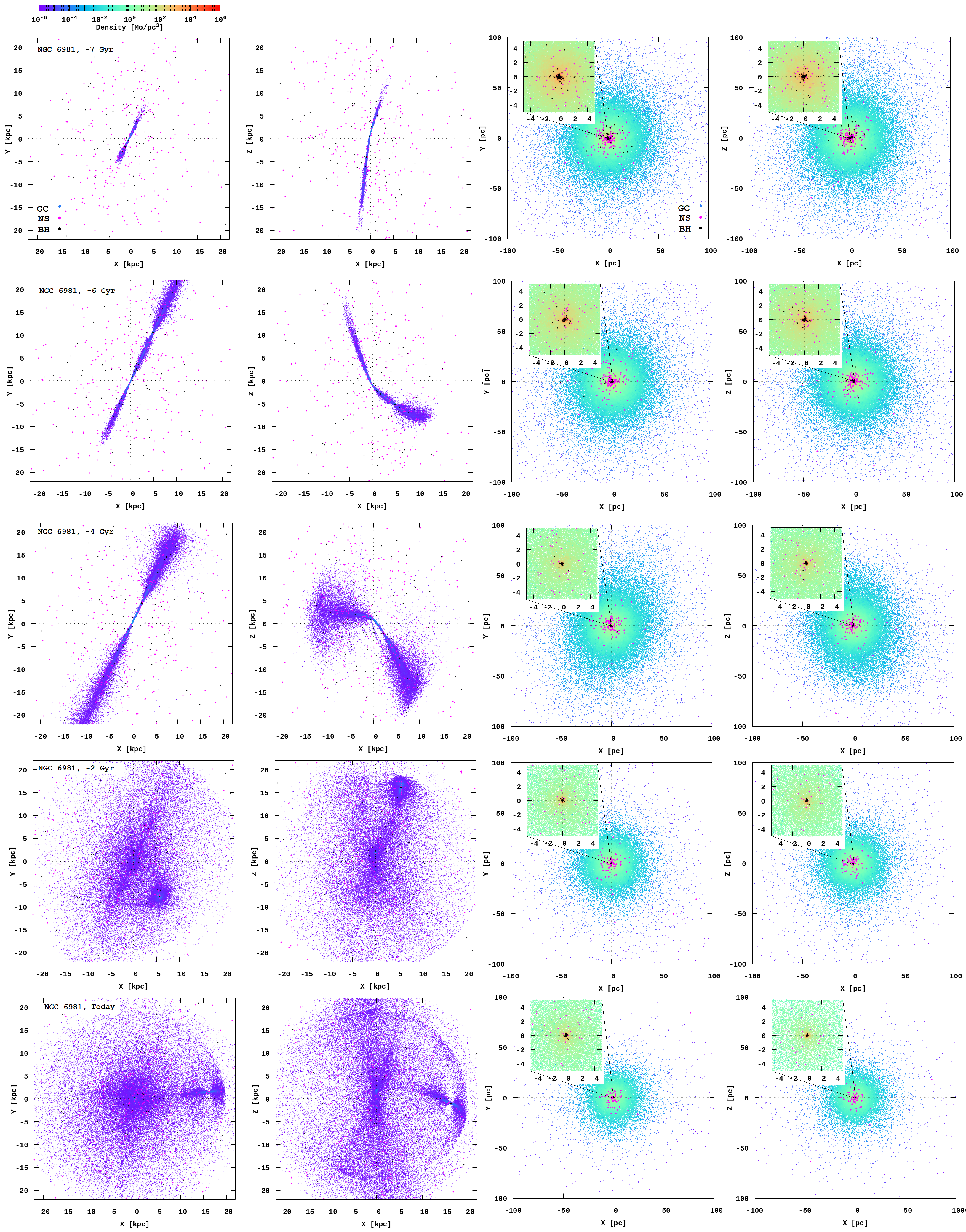}
\caption{Same as in Fig.~\ref{fig:ngc_6642_star_loss} for GC NGC~6981.}
\label{fig:ngc_6981_star_loss}
\end{figure*}

\begin{figure*}[ht]
\centering
\includegraphics[width=0.99\linewidth]{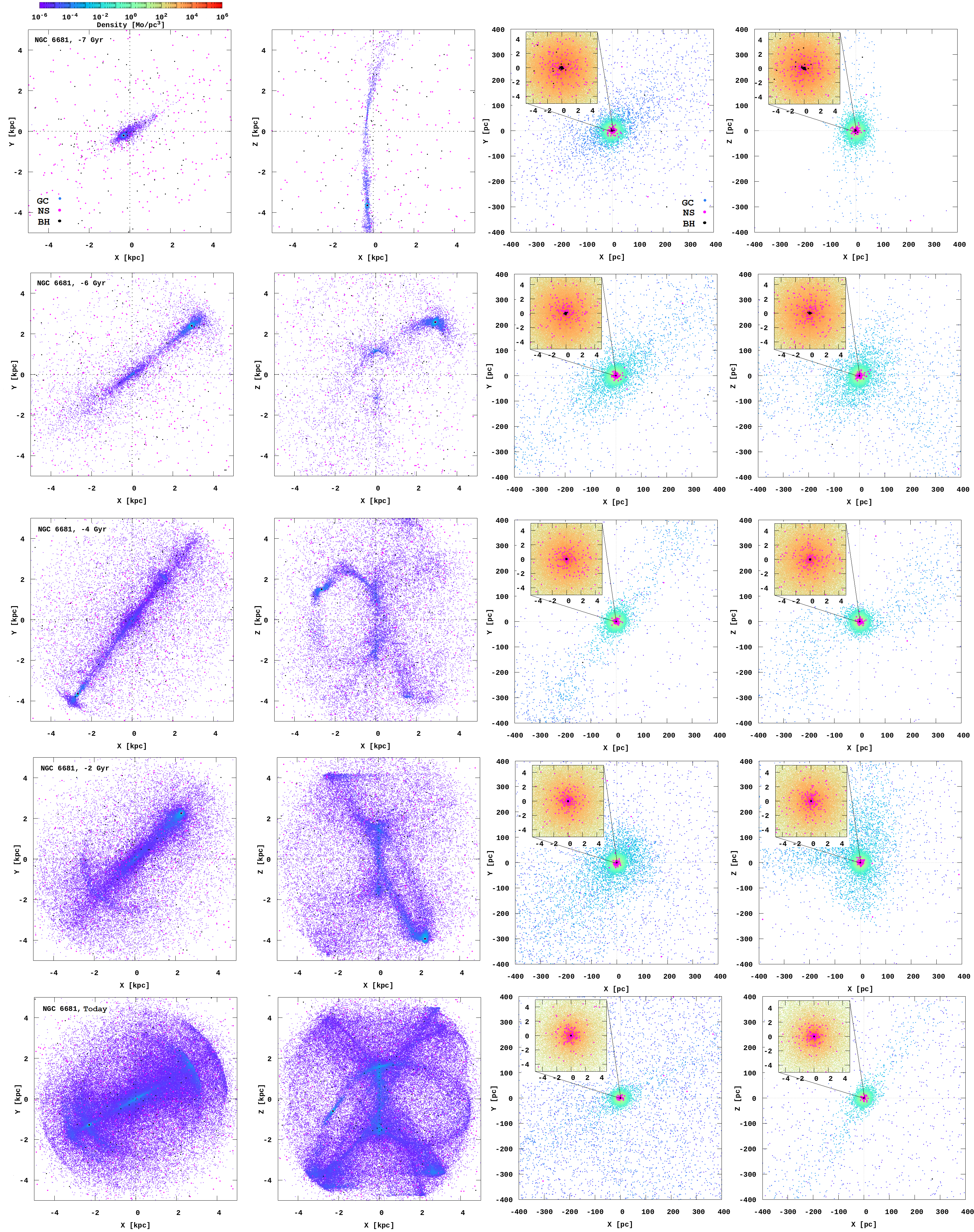}
\caption{Same as in Fig.~\ref{fig:ngc_6642_star_loss} for GC NGC~6681.}
\label{fig:ngc_6681_star_loss}
\end{figure*}

\begin{figure*}[ht]
\centering
\includegraphics[width=0.99\linewidth]{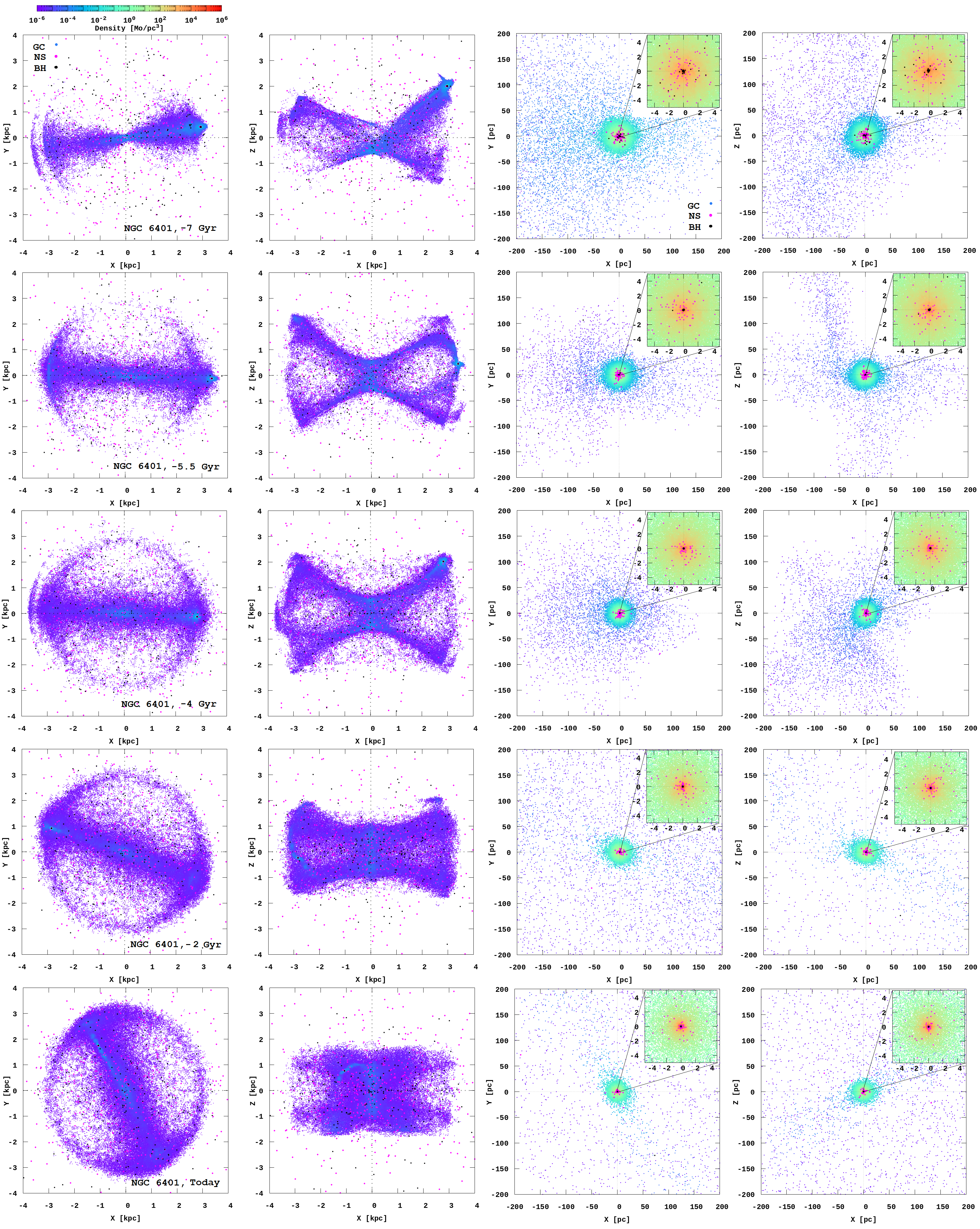}
\caption{Same as in Fig.~\ref{fig:ngc_6642_star_loss} for GC NGC~6401.}
\label{fig:ngc_6401_star_loss}
\end{figure*}

\begin{figure*}[ht]
\centering
\includegraphics[width=0.99\linewidth]{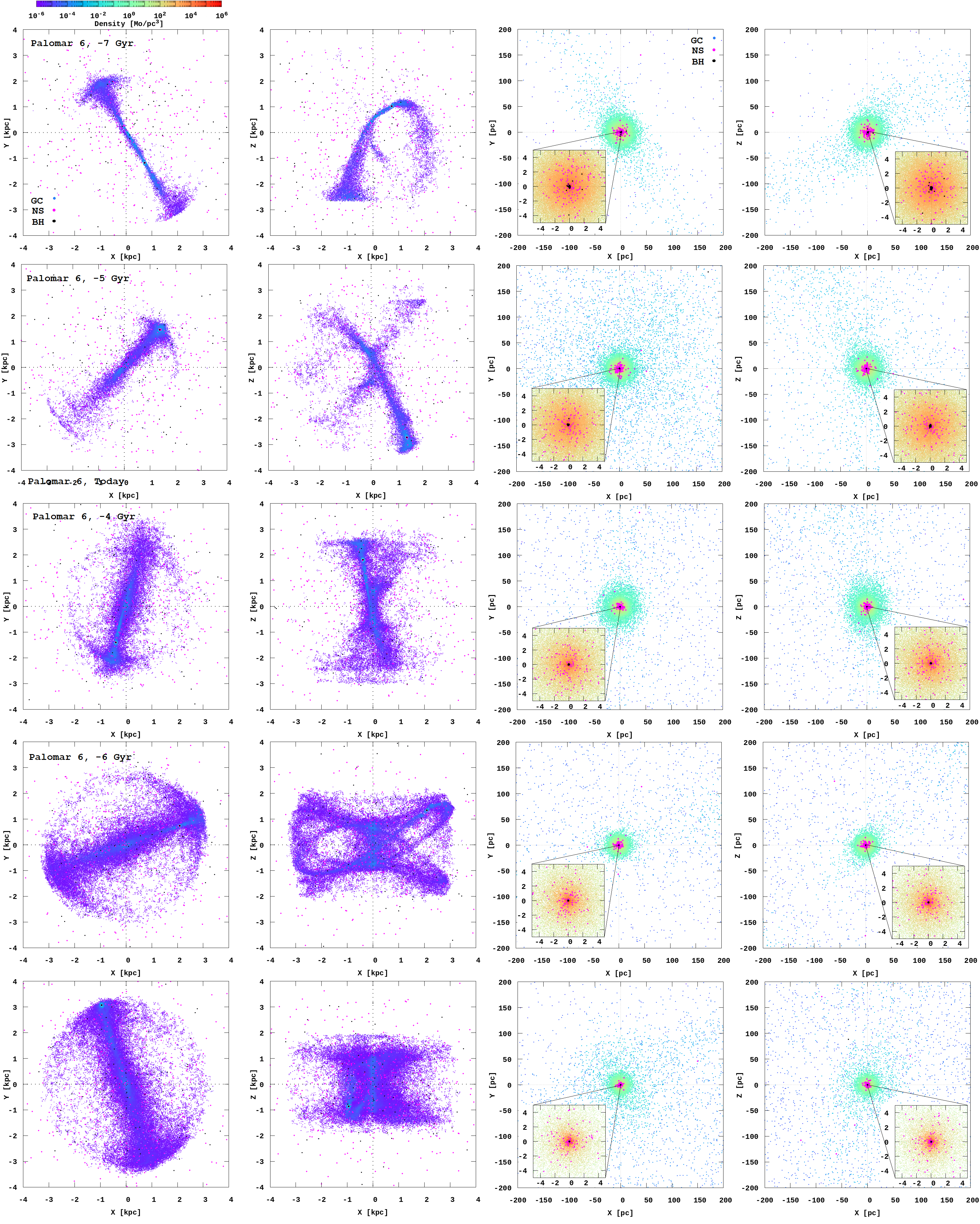}
\caption{Same as in Fig.~\ref{fig:ngc_6642_star_loss} for GC Palomar 6.}
\label{fig:ngc_pal6_star_loss}
\end{figure*}

\clearpage

\section{Total bound number of the stars with NSC for different velocity redaction factor}\label{app:bound-stars}

\begin{figure*}[htbp!]
\centering
\includegraphics[width=0.45\linewidth]{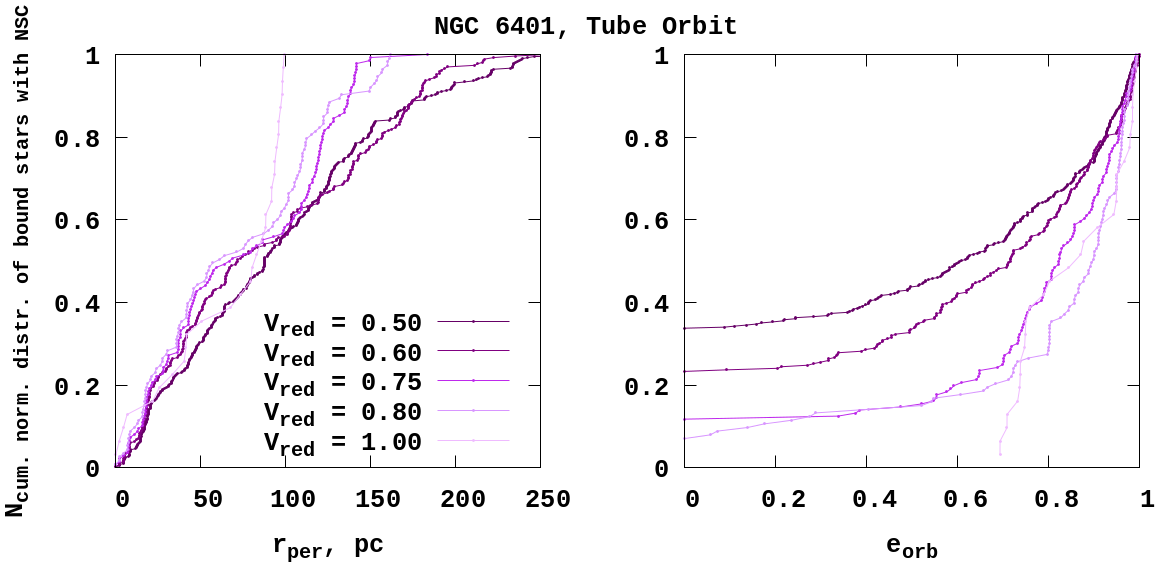}
\includegraphics[width=0.50\linewidth]{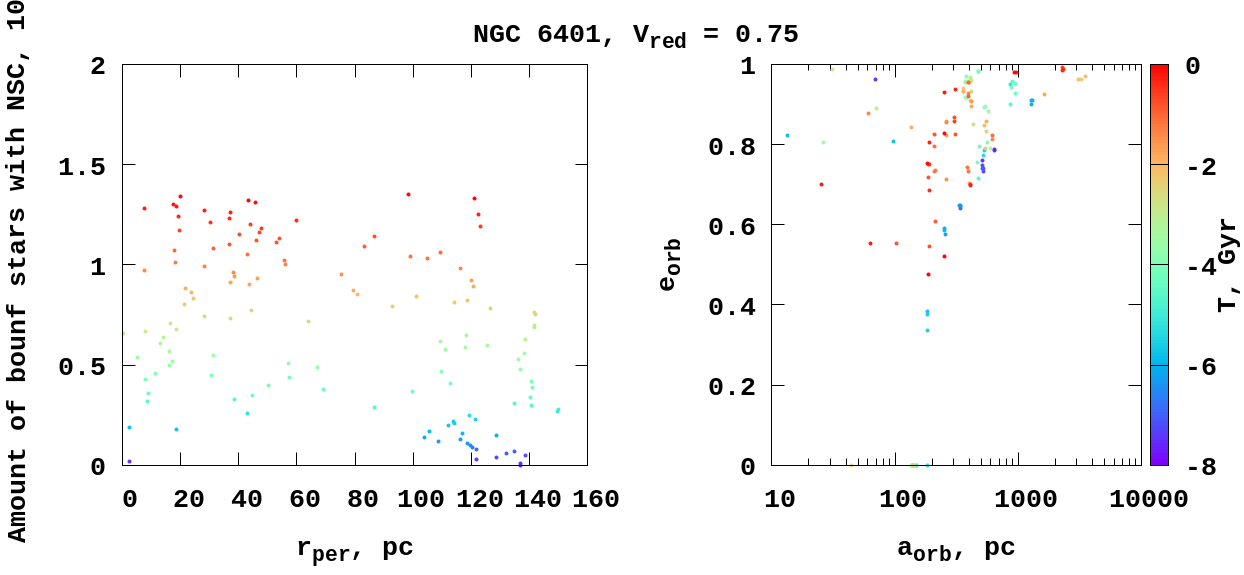}
\includegraphics[width=0.45\linewidth]{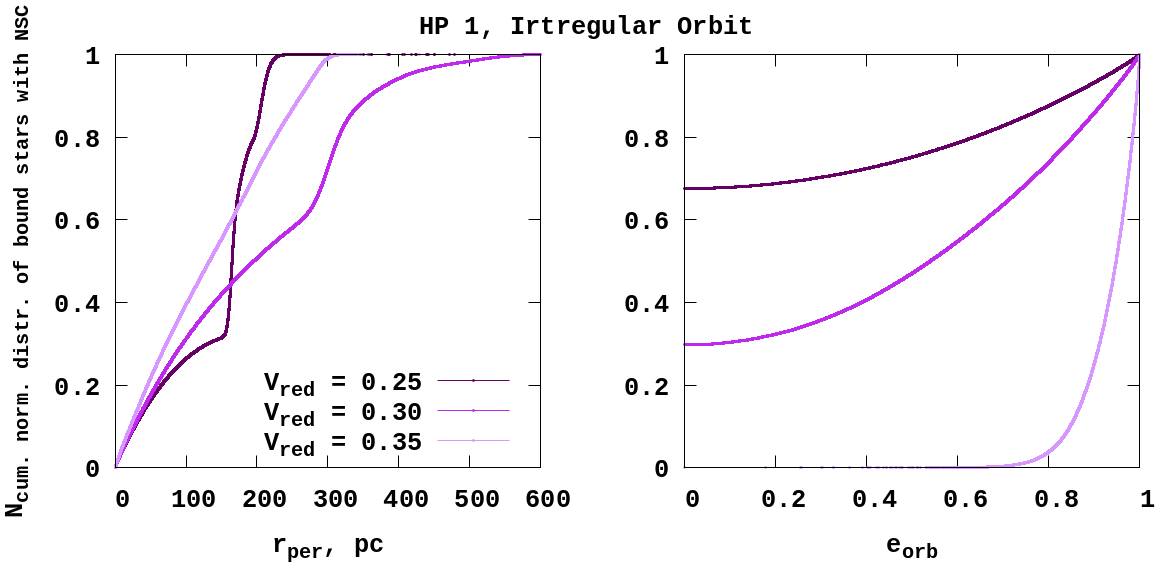}
\includegraphics[width=0.50\linewidth]{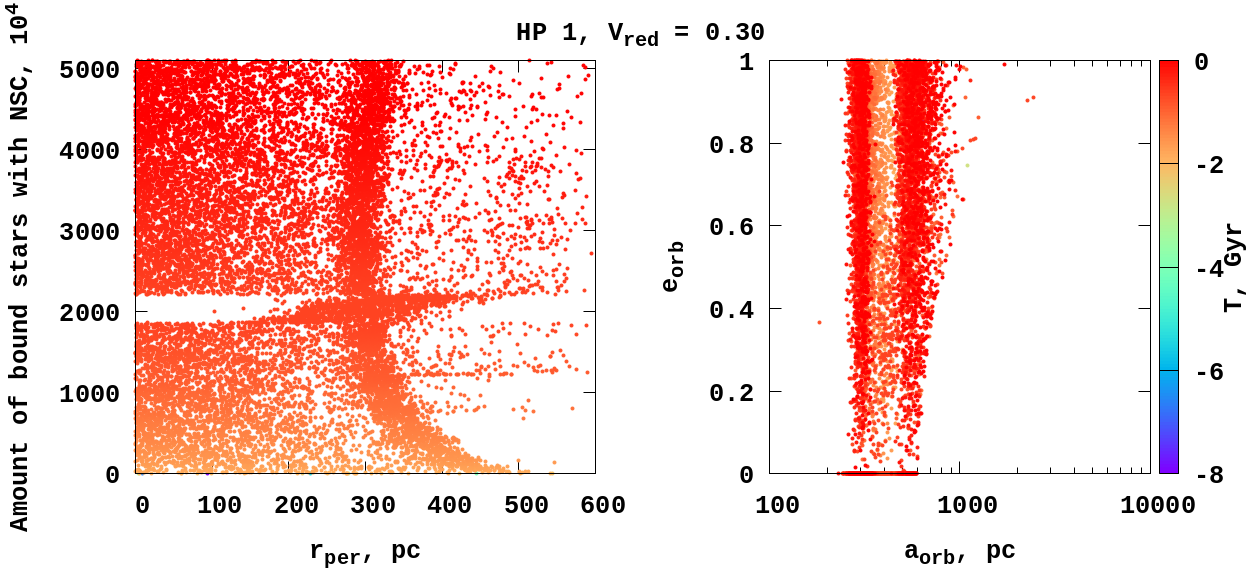}
\includegraphics[width=0.45\linewidth]{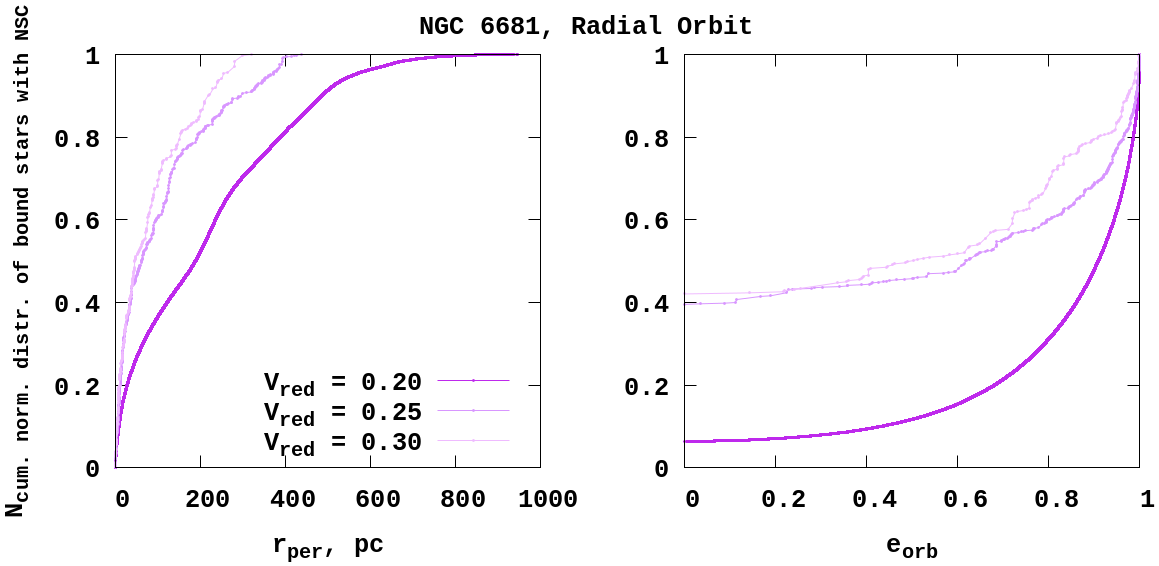}
\includegraphics[width=0.50\linewidth]{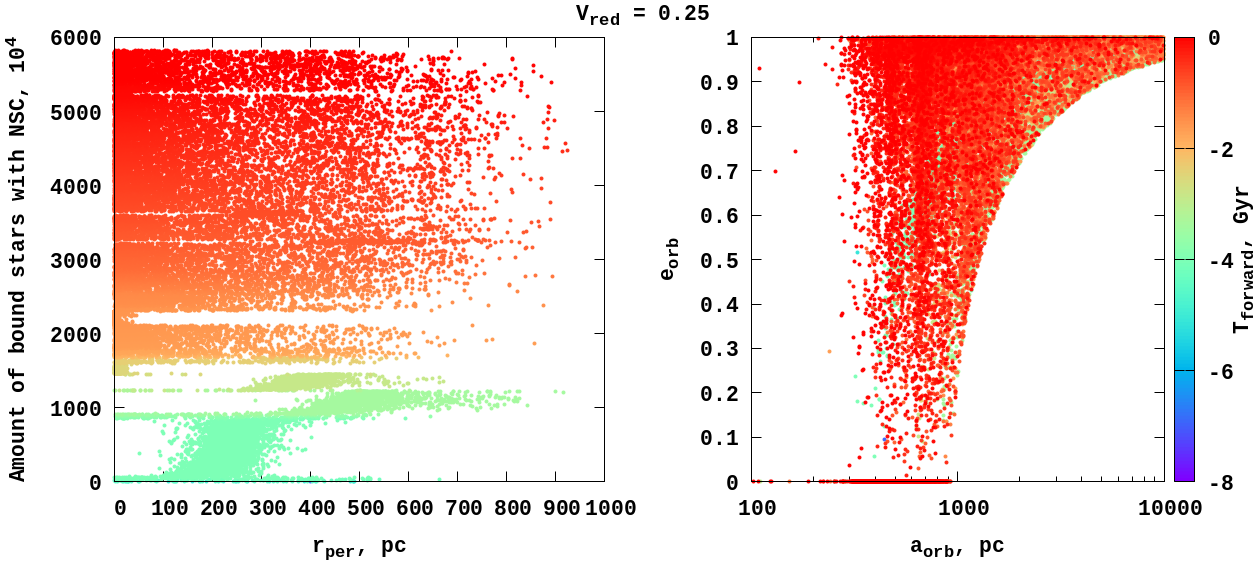}
\caption{Cumulative number distributions of the bound particles to the NSC as a function of the stars orbital parameters $r_{\rm per}$ and $e_{\rm orb}$ with different velocity reduction factors for NGC~6401, HP 1 and 6681. Dark violet -- V$_{\rm red}$ = 0.2; light violet -- V$_{\rm red}$ = 1.0, \textit{two left panels}. Total bound number of stars as a function of their percienter for different reduction $V_{\rm red}$.
Orbital elements e$_{\rm orb}$ vs. a$_{\rm orb}$ for the stars as a function of their GalC passage time, \textit{two right panels}.}
\label{fig:cum-red-gc}
\end{figure*}

\end{appendix}

\end{document}